\documentclass[aps,prb,twocolumn,showpacs,floatfix,superscriptaddress]{revtex4-1}

\usepackage[caption=false]{subfig}
\usepackage{verbatim}
\usepackage{graphicx}
\usepackage{dcolumn}
\usepackage{bm}
\usepackage{color}
\usepackage[colorlinks=true, allcolors=blue]{hyperref}
\usepackage{makeidx}
\usepackage{amsmath}
\usepackage{amssymb}
\usepackage{braket}
\usepackage{mathrsfs}
\makeindex

\DeclareMathOperator{\dif}{\mathrm{d}\!}
\renewcommand{\epsilon}{\varepsilon}

\begin{document}
\title{Probing interactions via nonequilibrium momentum distribution and noise in integer quantum Hall systems at $\nu=2$}
\author{Matteo Acciai}
\email{acciai@fisica.unige.it}
\affiliation{Dipartimento di Fisica, Universit\`a di Genova, Via Dodecaneso 33, 16146, Genova, Italy}
\affiliation{SPIN-CNR, Via Dodecaneso 33, 16146, Genova, Italy}
\affiliation{Aix Marseille Univ, Universit\'e de Toulon, CNRS, CPT, Marseille, France}
\author{Matteo Carrega}
\affiliation{NEST, Istituto Nanoscienze-CNR and Scuola Normale Superiore, Piazza San Silvestro 12, 56127 Pisa, Italy}
\author{J\'er\^ome Rech}
\affiliation{Aix Marseille Univ, Universit\'e de Toulon, CNRS, CPT, Marseille, France}
\author{Thibaut Jonckheere}
\affiliation{Aix Marseille Univ, Universit\'e de Toulon, CNRS, CPT, Marseille, France}
\author{Thierry Martin}
\affiliation{Aix Marseille Univ, Universit\'e de Toulon, CNRS, CPT, Marseille, France}
\author{Maura Sassetti}
\affiliation{Dipartimento di Fisica, Universit\`a di Genova, Via Dodecaneso 33, 16146, Genova, Italy}
\affiliation{SPIN-CNR, Via Dodecaneso 33, 16146, Genova, Italy}
\date{\today}
\begin{abstract}
We consider the excitation of single-electron wave packets by means of a time dependent voltage applied to the ballistic edge channels of the integer quantum Hall effect at filling factor $\nu=2$. Due to electron-electron interactions, fractional excitations emerge along the edge. Their detailed structure is analyzed by evaluating the non-equilibrium momentum distributions associated with the different edge channels. We provide results for a generic time-dependent drive both in the stationary regime and for intermediate times, where the overlap between fractionalized wave packets carries relevant information on interaction strength. As a particular example we focus on a Lorentzian drive, which provides a clear signature of the minimal excitations known as Levitons. Here, we argue that inner-channel fractionalized excitations can be exploited to extract information about inter-channel interactions. We further confirm this idea  by calculating the zero frequency noise due to the partitioning of these excitations at a quantum point contact and we propose a measurable quantity as a tool to directly probe electron-electron interactions and determine the so-called mixing angle of copropagating quantum Hall channels.
\end{abstract}

\maketitle
\section{Introduction}
Since its discovery, the quantum Hall (QH) effect\cite{klitzing80,girvin1990,girvin99,stern-review,goerbig09} has been the object of an intense research activity in condensed matter physics. This peculiar state of matter hosts ballistic one-dimensional chiral edge channels, where backscattering is forbidden by topological protection\cite{goerbig09,stern-review}. For this reason, quantum Hall-based systems have recently become of great interest for electron quantum optics (EQO)\cite{grenier11,bocquillon14}. This branch of mesoscopic physics is the condensed matter counterpart of traditional quantum optics, where electrons instead of photons are coherently manipulated at single-particle level. Here, quantum point contacts (QPCs) can be used as the electronic analog of beamsplitters, and chiral quantum Hall edge channels come as the most natural waveguides for electrons. In passing, we mention that also helical edge states of two-dimensional topological insulators\cite{haldane1988prl,konig07qsh,hasan2010colloquium} have been theoretically considered as an alternative platform \cite{hofer2013,inhofer2013,ferraro14HOMtopo,strom15entanglement,dolcini16chiral,calzona16energypart,acciai17}.

Recently, the fast development of EQO has been boosted thanks to the implementation of single-electron sources, like the driven mesoscopic capacitor\cite{feve07,moskalets08meso,mahe2010} and Lorentzian voltage pulses\cite{dubois2013levitonsNature,jullien14tomography}. In particular, the second approach has been shown to generate soliton-like states known as Levitons, theoretically proposed several years ago\cite{levitov96,levitov97}. Following these developments, several studies\cite{bocquillon12eqo,bocquillon13homscience,freulon15hom,tewari2016} have been realized, implementing the analog of Hanbury Brown-Twiss (HBT)\cite{hbt56} and Hong-Ou-Mandel (HOM)\cite{hom87} interferometers, allowing us to investigate statistical properties of excitations generated by the injection of few-electron wave packets. 

A major difference between EQO and conventional quantum optics is due to the fact that electrons interact with each other via screened Coulomb interactions.
It is well known that the latter are responsible for dramatic effects in one-dimensional (1D) channels in comparison to what happens in higher dimensions. Indeed, the low-energy physics of interacting 1D systems is  well-described by Luttinger liquid theory\cite{tomonaga,luttinger,haldane81,giamarchi}, replacing Landau's theory of Fermi liquid. The former predicts  intriguing phenomena such as spin-charge separation\cite{auslaender02,jompol-spincharge,hashisaka17spincharge} and fractionalization\cite{maslov-stone,safi95,safi97,pham2000,steinberg07chargefrac,kamata14,perfetto14timeresolved,calzona15physicaE,calzona15spin,karzig11,calzona17quench}. All these features originate from the fact that 1D systems have to be described in terms of collective excitations and QH edge states represent a formidable playground to inspect interaction effects.

Moreover, in recent years, non-equilibrium properties and interaction effects of quantum Hall edge states have been the object of an intense research activity on its own. Some examples are the study of interaction effects in Mach-Zehnder interferometers\cite{chalker2007MZ,levkivskyi08modelnu2,kovrizhin2009MZ,kovrizhin2010MZ} and interaction-induced decoherence of few-electron excitations\cite{degiovanni2009relaxation,degiovanni2010relaxation,jonckheere12hom,grenier13,ferraro14decoherence,wahl14prl,marguerite16decoherence,sukhorukov2016prb,cabart18}. In particular, the physics of QH systems at filling factor $\nu=2$, and the problem of equilibration when one of the edge channels is driven out of equilibrium due to a biased QPC has been considered both theoretically\cite{kovrizhin2011,kovrizhin2012,levkivskyi12relaxation} and experimentally\cite{sueur2010relaxation,altimiras2010relaxation,altimiras2010relaxtuning,itoh18-metastable-Hall,paradiso2011,paradiso2012prl,guiducci2018}. In particular, the non-equilibrium energy distribution of electrons has been measured\cite{sueur2010relaxation,altimiras2010relaxation,altimiras2010relaxtuning,itoh18-metastable-Hall} as a function of the propagation distance along the edge, showing that a non-thermal steady state is eventually reached\cite{itoh18-metastable-Hall}. All these studies demonstrate that dealing with interactions is essential to understand the physics of EQO in copropagating quantum Hall channels. However, a microscopic description of few-electron excitations relevant for EQO, as well as the possibility to use them as a probe of interactions, has received less attention to date.

This is the problem we address in the present paper. Here, we consider the excitation of single- or few-electron wave packets in the edge states of the integer quantum Hall effect at filling factor $\nu=2$, by using time-dependent voltage pulses. The presence of screened Coulomb interactions between copropagating channels leads to the fractionalization of the \emph{injected} pulse and create collective excitations propagating at different velocities and carrying an interaction-dependent fraction of the charge of the original wave packet\cite{levkivskyi08modelnu2,berg2009fractionalization,neder2012,milletari13,bocquillon2013,inoue2014}. 
In order to study their detailed structure, with a particular focus on particle-hole pair production, we evaluate non-equilibrium distributions in momentum space, which allow us to clearly distinguish, at a given momentum, between electron and hole contributions to a given excitation.

At first, we study the stationary regime where fractionalized excitations are well separated in space, so that they can be addressed independently. As a result, we show that non-equilibrium momentum distributions of each wave packet provide clear signatures of Levitons, i.e., minimal excitations generated by a Lorentzian drive.Then we focus on the inner channel and we study the transient regime taking into account the overlap between different excitations. In this regime, it is shown that the overlap contribution to the momentum distribution results in the reduction of the overall number of electron and hole excitations with respect to the stationary one. Based on the previous findings, we suggest that inner-channel excitations, instead of the more addressed outer ones, could be exploited to extract the strength of inter-channel interactions.

We further develop this idea by considering the noise generated by partitioning these excitations at a QPC. In particular, we will show that it is possible to construct a measurable quantity, based on the expected noise behavior, demonstrating that, for a periodic train of rectangular pulses, it presents peculiar features which enable to directly determine the mixing angle. Finally, by exploiting the experimentally relevant case of a periodic voltage drive, we propose a tool to directly access and probe electron-electron interactions.

The paper is organized as follows: in Sec.\ \ref{sec:model} we introduce the model and the setup; Sec.\ \ref{sec:nk} presents the calculation of the distributions of fractional excitations in momentum space; in Sec.\ \ref{sec:noise} we discuss the noise produced by partitioning fractional excitations at a QPC. Finally, Sec.\ \ref{sec:conclusions} contains our conclusions.

We choose units such that $\hbar=k_\text{B}=1$.

\section{Model and setup}\label{sec:model}
\begin{figure}[t]
	\centering
	\includegraphics[width=\columnwidth]{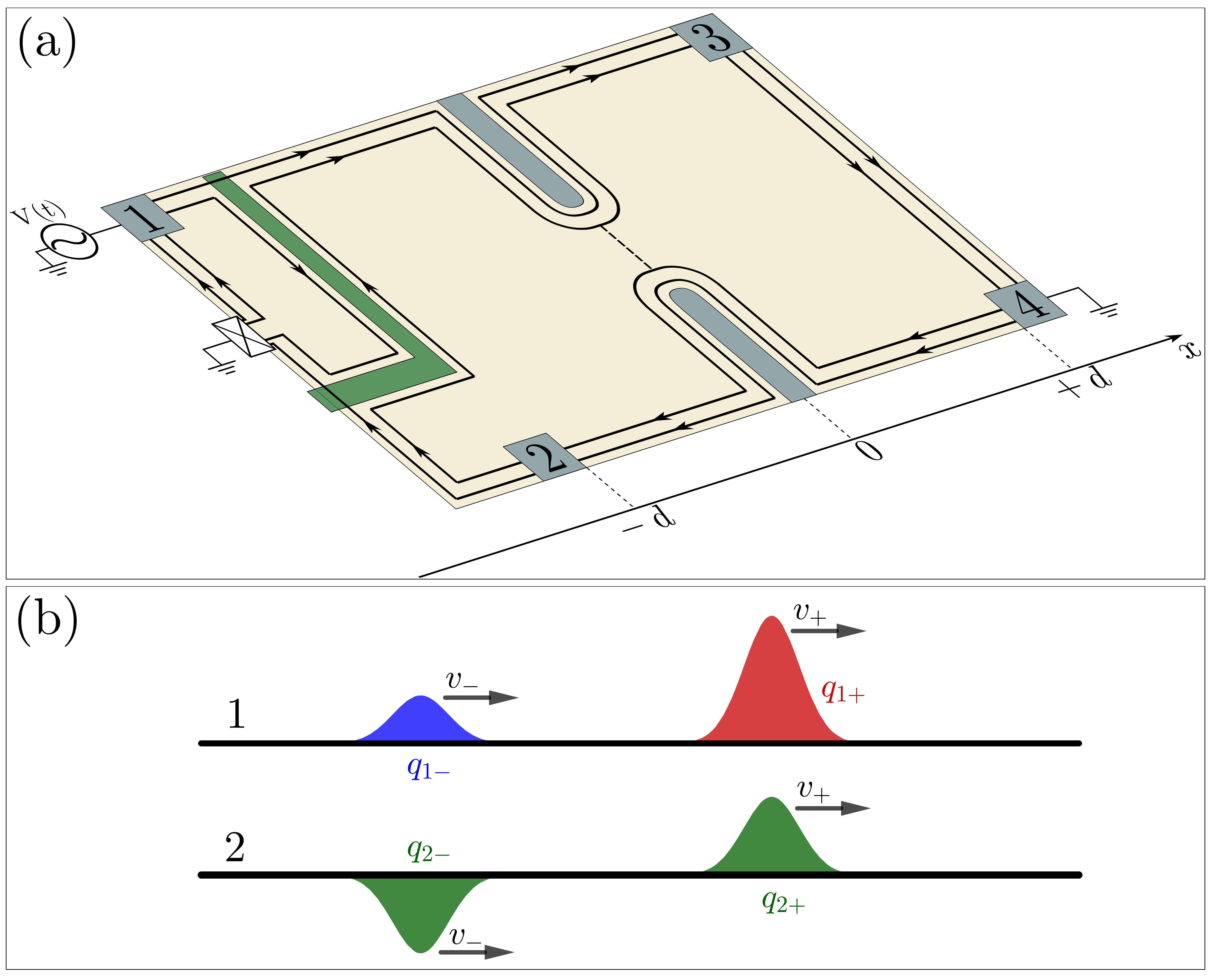}
	\caption{(a) Sketch of the considered setup. A time-dependent external voltage drive applied at terminal 1 drives the system out of equilibrium, exciting electron wave packets into the edge channels. Right after the driven contact, a gate (green) acts as a filter, transmitting only outer-channel excitations. After a propagation distance $d$, a quantum point contact is set to fully transmit the outer channel and partially reflect the inner one, thus creating current fluctuations at terminals 2 and 3 (see Sec.\ \ref{sec:noise}). (b) Sketch of the fractionalization of the excited pulse: for $q>0$ [see Eq.\ \eqref{eq:definition-q}] two electron-like excitations are created in the outer channel, while hole-like $(q_{2-})$ and electron-like $(q_{2+})$ excitations emerge in the inner channel, to which it is not transferred any net charge.}
	\label{fig:setup}
\end{figure}

We consider a quantum Hall bar at filling factor $\nu=2$. In this regime, two copropagating ballistic channels arise at each edge of the system, as depicted in Fig.\ \ref{fig:setup}(a). We will use index $\alpha=1,2$ to label the outer and inner channels, respectively; indices $r=R(=+)$ and $r=L(=-)$ will denote the right and left moving edges.
\subsection{Hamiltonian and equations of motion}
 The Hamiltonian of the edge channels is $H_\text{edge}=H_0+H_\text{int}$, where
\begin{equation}
H_0=\sum_{r=R,L}\sum_{\alpha=1,2}v_\alpha\int dx\,\Psi^\dagger_{r\alpha}(x)(-ir\partial_x)\Psi_{r\alpha}(x)
\end{equation}
is the kinetic term and
\begin{equation}
H_\text{int}=2\pi u\sum_{r=R,L}\int dx\, n_{r1}(x)n_{r2}(x)
\end{equation}
describes short range Coulomb interactions, with coupling strength $u$, between different channels along a given edge $r=R,L$. For sake of simplicity, intra-channel interactions are not considered here, since their effects can be taken into account by properly renormalizing the edge velocities $v_\alpha$. It is worth mentioning that the above model for short range Coulomb interactions\cite{haldane81,levkivskyi08modelnu2} has been successfully used to interpret different experiments and is in good agreement with very recent measurements of non-equilibrium energy distributions in a $\nu=2$ system\cite{itoh18-metastable-Hall}. Operators $\Psi_{r\alpha}$ appearing in $H_0$ are fermionic fields, satisfying anticommutation relations $\{\Psi_{r\alpha}(x),\Psi^\dagger_{r'\alpha'}(x')\}=\delta_{r,r'}\delta_{\alpha,\alpha'}\delta(x-x')$ and $n_{r\alpha}(x)=\Psi^\dagger_{r\alpha}(x)\Psi_{r\alpha}(x)$ are particle density operators.

To deal with Coulomb interactions, fermion operators can be expressed in terms of bosonic ones through the standard bosonization procedure\cite{vondelft}
\begin{equation}
\Psi_{r\alpha}(x)=\frac{\mathcal{F}_{r\alpha}}{\sqrt{2\pi a}}e^{-i\sqrt{2\pi}\Phi_{r\alpha}(x)}\,,
\label{eq:bosonization}
\end{equation}
where $a$ is the usual short distance cutoff and $\mathcal{F}_{r\alpha}$ are Klein factors\cite{vondelft,safi02klein,meden2016}. Bosonic fields $\Phi_{r\alpha}$ satisfy the commutation relations $2[\Phi_{r\alpha}(x),\Phi_{r'\alpha'}(x')]=ir\delta_{r,r'}\delta_{\alpha,\alpha'}\mathrm{sgn}(x-x')$ and are related to the particle density operators by $n_{r\alpha}(x)=-r(2\pi)^{-1/2}\partial_x\Phi_{r\alpha}(x)$. Thanks to this and to the fact that $\int dx\,2\Psi^\dagger_{r\alpha}(x)(-ir\partial_x)\Psi_{r\alpha}(x)=\int dx[\partial_x\Phi_{r\alpha}(x)]^2$, the Hamiltonian $H_\text{edge}$ is quadratic when expressed in terms of bosonic field operators.
Then, the rotation
\begin{equation}
\begin{pmatrix}
\Phi_{r1}\\
\Phi_{r2}\\
\end{pmatrix}=
\begin{pmatrix}
\cos\theta & -\sin\theta\\
\sin\theta & \cos\theta\\
\end{pmatrix}
\begin{pmatrix}
\Phi_{r+}\\
\Phi_{r-}\\
\end{pmatrix},\,\,\tan2\theta=\frac{2u}{v_1-v_2}
\label{eq:diag}
\end{equation}
brings the edge Hamiltonian in diagonal form:
\begin{equation}
H_\text{edge}=\sum_{r=R,L}\sum_{\eta=\pm}\frac{v_\eta}{2}\int dx[\partial_x\Phi_{r\eta}(x)]^2\,.
\end{equation}
Here, $\Phi_{r\pm}$ are chiral fields (left- or right-moving depending on the edge index $r$), describing free excitations (called fast and slow modes in the literature) propagating at velocities $v_\pm=(v_1+v_2)/2\pm u/\sin2\theta$, in the same direction on a given edge.

Next, we consider a time-dependent voltage drive, as sketched in Fig.\ \ref{fig:setup}(a), which brings the system out-of-equilibrium, generating electron wave packets on the edge channels\cite{dubois2013levitonsNature,dubois13prb,rech16prl,vannucci17heat}. The excitation of wave packets along the upper-edge outer channel is modeled by the Hamiltonian
\begin{equation}
H_\text{V}=-e\int dx\, U(x,t)n_{R1}(x)\,,
\label{eq:source}
\end{equation}
$-e$ being the electron charge and $U(x,t)$ an external voltage drive.
In the following we shall consider
\begin{equation}
U(x,t)=\Theta(-x-d)V(t)\,,
\label{eq:gate}
\end{equation}
with $\Theta(x)$ the Heaviside step function, to describe that the voltage drive $V(t)$ is applied at terminal $1$, i.e.\ for $x<-d$ [see Fig.\ \ref{fig:setup}(a)].
A practical way to create a wave packet only in the outer channel could be to use a gate as a filter right after the driven contact, similarly to what has been done in Ref.\ \onlinecite{hashisaka17spincharge}. When the gate is biased so that the outer channel is fully transmitted and the inner one is fully reflected, the desired injection on the outer channel is achieved, while no excitation is transmitted to the inner channel.
This situation can be modeled as if the external drive $U(x,t)$ only couples to the outer-channel charge density $n_{R1}(x)$, as we considered in \eqref{eq:source}. The equations of motion in presence of the coupling to the external drive read
\begin{equation}
(\partial_t+v_\pm\partial_x)\Phi_{r\pm}(x,t)=\frac{\zeta_{r,\pm}}{\sqrt{2\pi}}U(x,t)\,,
\end{equation}
with $\zeta_{L,\pm}=0$, $\zeta_{R,+}=-e\cos\theta$ and $\zeta_{R,-}=e\sin\theta$. These are solved by\cite{dolcini16chiral,rech16prl,ronetti17polarized}
\begin{equation}
\Phi_{r\pm}(x,t)=\phi_{r,\pm}(x-v_\pm t,0)+\frac{\zeta_{r\pm}}{\sqrt{2\pi}}\varphi_\pm(x,t)\,,
\label{eq:solution-eom-1}
\end{equation}
where $\phi_{r\pm}(x,t)$ are the free bosonic fields in the absence of the drive and
\begin{equation}
\varphi_\pm(x,t)=\int_{-\infty}^{t}dt'\,U[x-v_\pm(t-t'),t']
\label{eq:solution-eom-2}
\end{equation}
describe the effect of the applied voltage. Similarly, the time evolution of fermionic operators can be written as
\begin{equation}
\Psi_{r\alpha}(x,t)=\psi_{r\alpha}(x,t)e^{-\frac{i}{e}\sum_{\eta=\pm}\sigma_{r\alpha\eta}\varphi_\eta(x,t)}\,,
\label{eq:evolution-Psi}
\end{equation}
with $\psi_{r\alpha}(x,t)$ describing the time evolution when the external drive is absent and $\sigma_{r1\eta}=-\zeta_{r,\eta}^2$, $\sigma_{r2\eta}=\eta\zeta_{r,\eta}\zeta_{r,-\eta}$.
This solution holds true for an arbitrary voltage drive, with the only constraint that $\lim_{t\to-\infty}U(x,t)=0$.

Equations \eqref{eq:solution-eom-1} and \eqref{eq:solution-eom-2}, together with \eqref{eq:bosonization} and \eqref{eq:diag}, completely determine the dynamics of the system and enable one to compute expectation values of bosonic as well as fermionic correlators.
We shall denote these averages with the symbol $\Braket{\dots}$; since the whole time evolution is attributed to operators, thermal averages are computed with respect to the initial equilibrium density matrix at $t=-\infty$, characterized by temperature $T$.

\subsection{Charge fractionalization}
We now briefly recall some aspects of charge fractionalization occurring in interacting 1D channels\cite{levkivskyi08modelnu2,kovrizhin2012,milletari13,inoue2014}. We remind the reader that, due to Coulomb interactions, an electron wave packet generated in a 1D channel splits up into fractionalized wave packets carrying fractions of the electron charge.
Here, we compute the excess charge densities $\Delta\rho_{r\alpha}(x,t)$, i.e.\ the deviation of the expectation values of particle density operators $\rho_{r,\alpha}(x,t)=-en_{r\alpha(x,t)}$ with respect to the equilibrium situation where no external drive is applied. The result is $\Delta\rho_{L\alpha}(x,t)=0$ and
\begin{align}
\Delta \rho_{R1}(x,t)&=\frac{e^2}{2\pi}\Theta(x+d)\left[\frac{\sin^2\!\theta}{v_-}V\left(t-\frac{x+d}{v_-}\right)\right.\notag\\
&+\left.\frac{\cos^2\!\theta}{v_+}V\left(t-\frac{x+d}{v_+}\right)\right]\,,\\
\Delta \rho_{R2}(x,t)&=\frac{e^2\sin 2\theta}{4\pi}\Theta(x+d)\left[-\frac{1}{v_-}V\left(t-\frac{x+d}{v_-}\right)\right.\notag\\
&+\left.\frac{1}{v_+}V\left(t-\frac{x+d}{v_+}\right)\right]\,.
\end{align}
This shows that fast and slow excitations, retaining the shape of the applied voltage pulse and propagating at velocities $v_\pm$ respectively, emerge in both the outer and inner channels. The total charge carried by the fast/slow excitation on channel $\alpha$ is given by $-eq_{\alpha\pm}$, with
\begin{equation}
\begin{gathered}
q_{1+}=q\cos^2\!\theta\,,\quad q_{1-}=q\sin^2\!\theta\,,\\
q_{2\pm}=\pm q\cos\theta\sin\theta\equiv\pm q_2
\end{gathered}
\label{eq:frac-charges}
\end{equation}
where the dimensionless parameter $q$ is defined as
\begin{equation}
q=-\frac{e}{2\pi}\int_{-\infty}^{+\infty}dt\,V(t)\,.
\label{eq:definition-q}
\end{equation}
Notice that this quantity is related only to the external voltage and represents the total excess charge (in units of $-e$) present on the outer channel due to the coupling to $V(t)$. We stress that no net charge is transferred to the inner channel, since its fractional excitations have equal and opposite charges. A sketch of the fractionalization is represented in Fig.\ \ref{fig:setup}(b). For the sake of definiteness, we will only consider the case $q>0$ (electron-like pulse).
In the next section we will analyze the properties of these fractional excitations in detail by computing their non-equilibrium momentum distribution.

\section{Momentum distribution}\label{sec:nk}
The above description of fractionalization relies on integrated quantities (i.e.\ charges) and, as such, does not give any direct information on the detailed structure of fractionalized excitations. In particular, it is not sensitive to their particle-hole content. In order to investigate these features, we can resort to the non-equilibrium momentum distribution, which allows for a microscopic investigation of non-equilibrium dynamics and interaction effects, as we will show below.
We will consider only the dynamics of the upper edge, where the voltage is applied and fractional pulses are generated. Throughout this section we thus drop the index $r$, since we will only refer to $r=R$.

Non-equilibrium momentum distributions are defined as follows:
\begin{equation}
\Delta n_\alpha(k,t)=\Braket{c_\alpha^\dagger(k,t)c_\alpha(k,t)}-n_\alpha^{(0)}(k)\,,
\end{equation}
with $c_\alpha(k)$ the operator annihilating an electron with momentum $k$ on channel $\alpha$. Note that we focus again on the variation with respect to the initial equilibrium distribution $n_\alpha^{(0)}(k)$. As a consequence, positive (negative) values of $\Delta n_{\alpha}$ will describe electron (hole) excitations. Moreover, since the external voltage drives the system out of equilibrium, the momentum distribution acquires, in general, a non-trivial time dependence.

The previous expression can be conveniently written as (we set the Fermi momentum to $k_\text{F}=0$)
\begin{equation}
\Delta n_\alpha(k)=\frac{1}{2\pi}\int_{-\infty}^{+\infty}dx\int_{-\infty}^{+\infty}d\xi\,e^{-ik\xi}\Delta\mathscr{G}_\alpha(x,\xi;t)\,,
\label{eq:nk}
\end{equation} 
where we have introduced the equal-time correlators
\begin{equation}
\begin{aligned}
\Delta\mathscr{G}_\alpha(x,\xi;t)&=\Braket{\Psi^\dagger_\alpha\left(x-\frac{\xi}{2},t\right)\Psi_\alpha\left(x+\frac{\xi}{2},t\right)}+\\
&\quad -\Braket{\psi^\dagger_\alpha\left(x-\frac{\xi}{2},t\right)\psi_\alpha\left(x+\frac{\xi}{2},t\right)}\,.
\end{aligned}
\end{equation}
They are computed with the help of standard bosonization techniques\cite{vondelft} and read
\begin{subequations}
\begin{equation}
\begin{aligned}
&\Delta\mathscr{G}_1(x,\xi;t)=\frac{1}{2\pi a}\,{e^{\cos^2\!\theta\,G_+(-\xi)+\sin^2\!\theta\,G_-(-\xi)}}\\
&\quad\times\left\{e^{i[\zeta_+\cos\theta\Delta\varphi_+(x,\xi;t)-\zeta_-\sin\theta\Delta\varphi_-(x,\xi;t)]}-1\right\}\,,
\end{aligned}
\end{equation}
\begin{equation}
\begin{aligned}
&\Delta\mathscr{G}_2(x,\xi;t)=\frac{1}{2\pi a}\,e^{\sin^2\!\theta\,G_+(-\xi)+\cos^2\!\theta\,G_-(-\xi)}\\
&\quad\times\left\{e^{i[\zeta_+\sin\theta\Delta\varphi_+(x,\xi;t)+\zeta_-\cos\theta\Delta\varphi_-(x,\xi;t)]}-1\right\}\,.\label{eq:correlator-et-2}
\end{aligned}
\end{equation}
\label{eq:correlators-et}
\end{subequations}
Here, we have introduced the equilibrium bosonic Green functions
\begin{equation}
\begin{aligned}
G_\pm(x)&=\Braket{\phi_\pm(x,0)\phi_\pm(0,0)-\phi_\pm^2(0,0)}\\
&=\ln\left[\frac{a}{a-ix}\frac{\pi Tx/v_\pm}{\sinh(\pi Tx/v_\pm)}\right]
\end{aligned}
\end{equation}
and the phase differences
\begin{equation}
\Delta\varphi_\pm(x,\xi;t)=\varphi_\pm\left(x-\frac{\xi}{2},t\right)-\varphi_\pm\left(x+\frac{\xi}{2},t\right)\,.
\end{equation}

For the moment, we assume the time-dependent voltage $V(t)$ to be a single pulse localized around $t=0$, with a characteristic temporal extension $w$. Moreover, we shall study the evolution of the system at times such that the external voltage pulse is negligible, i.e. $t\gtrsim w$. 
\subsection{Stationary regime}\label{sec:nk-stationary}
Because of the different propagation velocities $(v_+>v_-)$, fractional pulses on both the outer and inner channels become more and more separated in space as they propagate along the edge. We now consider the case when they are very well separated; since the fast/slow excitation are centered around $x_\pm=v_\pm t$ with spatial extension $\delta x_\pm=wv_\pm$, this is achieved when $|x_+-x_-|\gg|\delta x_++\delta x_-|$, i.e.\ for times $t$ such that $t\gg w\frac{v_++v_-}{v_+-v_-}$. As proved in Refs.\ \onlinecite{acciai17,sukhorukov2016prb}, in this regime it is possible to separate the correlators in Eq.\ \eqref{eq:correlators-et} as
\begin{equation} 
\Delta\mathscr{G}_\alpha(x,\xi;t)=\Delta\mathscr{G}_{\alpha+}(x_+,\xi)+\Delta\mathscr{G}_{\alpha-}(x_-,\xi)\,,
\label{eq:separation}
\end{equation}
where $x_\pm=x-v_\pm t$ and
\begin{subequations}
\begin{equation}
\begin{aligned}
\Delta\mathscr{G}_{\alpha+}(x_+,\xi)&=\frac{e^{s_{\alpha,+}G_+(-\xi)+s_{\alpha,-}\,G_-(-\xi)}}{2\pi a}\\
&\times\left[e^{i\zeta_+\partial_\theta^{2-\alpha}\sin\theta\,\Delta\varphi_+(x_+,\xi)}-1\right]\,,
\end{aligned}
\end{equation}
\begin{equation}
\begin{aligned}
\Delta\mathscr{G}_{\alpha-}(x_-,\xi)&=\frac{e^{s_{\alpha,+}G_+(-\xi)+s_{\alpha,-}G_-(-\xi)}}{2\pi a}\\
&\times\left[e^{i\zeta_-\partial_\theta^{2-\alpha}\cos\theta\,\Delta\varphi_-(x_-,\xi)}-1\right]\,,
\end{aligned}
\end{equation}
\label{eq:separated-correlators}
\end{subequations}
with $s_{1,+}=\cos^2\!\theta=s_{2,-}$ and $s_{1,-}=\sin^2\!\theta=s_{2,+}$.
Technical details on the aforementioned procedure can be found in Ref.\ \onlinecite{acciai17}. Here, we simply stress that the separation \eqref{eq:separation} means that fast and slow excitations on a given channel $\alpha$ are treated independently from one another, which is reasonable when their overlap in space is negligible. Now, thanks to Eqs.\ \eqref{eq:nk} and \eqref{eq:separation}, we have
\begin{equation}
\Delta n_\alpha(k,t)\to\Delta n_{\alpha+}(k)+\Delta n_{\alpha-}(k)\,.
\end{equation}
Here, distributions on the r.h.s.\ of the last equation are given by \eqref{eq:nk}, with correlators \eqref{eq:separated-correlators} instead of \eqref{eq:correlators-et}, and are time-independent thanks to the fact that $\Delta\mathscr{G}_{\alpha\pm}$ depend on space and time only via the combinations $x-v_\pm t$\cite{acciai17}. Therefore any dependence on $t$ is lost when computing the integral over $x$ in \eqref{eq:nk}. This is why we called this regime stationary.

We consider the particularly relevant case of a Lorentzian pulse
\begin{equation}
V(t)=-\frac{q}{e}\frac{2w}{w^2+t^2}\,,
\label{eq:lor-pulse}
\end{equation}
which, for positive integer $q$, is known to generate pure electron-like excitations without any particle-hole pair (Levitons) in non-interacting quantum conductors\cite{levitov96,levitov97,keeling06,dubois2013levitonsNature}, as well as in the strongly correlated fractional quantum Hall phase \cite{rech16prl,ronetti17levitons}.
For this drive the correlators have the following zero-temperature expression:
\begin{equation}
\begin{aligned}
&\Delta\mathscr{G}_{\alpha\pm}(x,\xi;t)=\frac{1}{2\pi i \xi}\times\\
&\left\{\prod_{\eta=\pm}\left[\frac{iv_\pm w-\eta(v_\pm t_\pm+\eta\xi/2)}{iv_\pm w+\eta(v_\pm t_\pm+\eta\xi/2)}\right]^{q_{\alpha\pm}}-1\right\}\,,
\end{aligned}
\label{eq:corr-lor}
\end{equation}
with $q_{\alpha\pm}$ given in \eqref{eq:frac-charges} and $t_\pm=t-x/v_\pm$. When $q_{\alpha\pm}$ are integer numbers, correlators $\Delta\mathscr{G}_{\alpha\pm}$ are thus analytic functions of $\xi$ (in the upper/lower half-plane, depending whether $q_{\alpha\pm}$ is positive/negative). Accordingly, when calculating the integral over $\xi$ to obtain the distributions $\Delta n_{\alpha\pm}$ [see Eq.\ \eqref{eq:nk}], the result vanishes for negative/positive momenta. Explicitly, for positive integer $q_{1+}=m_+$, $q_{1-}=m_-$ and $q_{2\pm}=\pm n$, we have the following expressions for the momentum distributions at $T=0$ (see Appendix\ \ref{app:nk-int}):
\begin{equation}
\begin{aligned}
\Delta n_{1\pm}(k)&=2wv_\pm\Theta(k)\sum_{j=0}^{m_\pm-1}\left|L_j(2wv_\pm k)\right|^2e^{-2wv_\pm k}\,,\\
\Delta n_{2\pm}(k)&=\pm 2wv_\pm\Theta(\pm k)\sum_{j=0}^{n-1}\left|L_j(\pm 2wv_\pm k)\right|^2e^{\mp 2wv_\pm k}\,,
\end{aligned}
\label{eq:nk-stationary}
\end{equation}
where $L_j$ are Laguerre polynomials.
\begin{figure}[tbp]
	\begin{center}
		\includegraphics[width=\columnwidth]{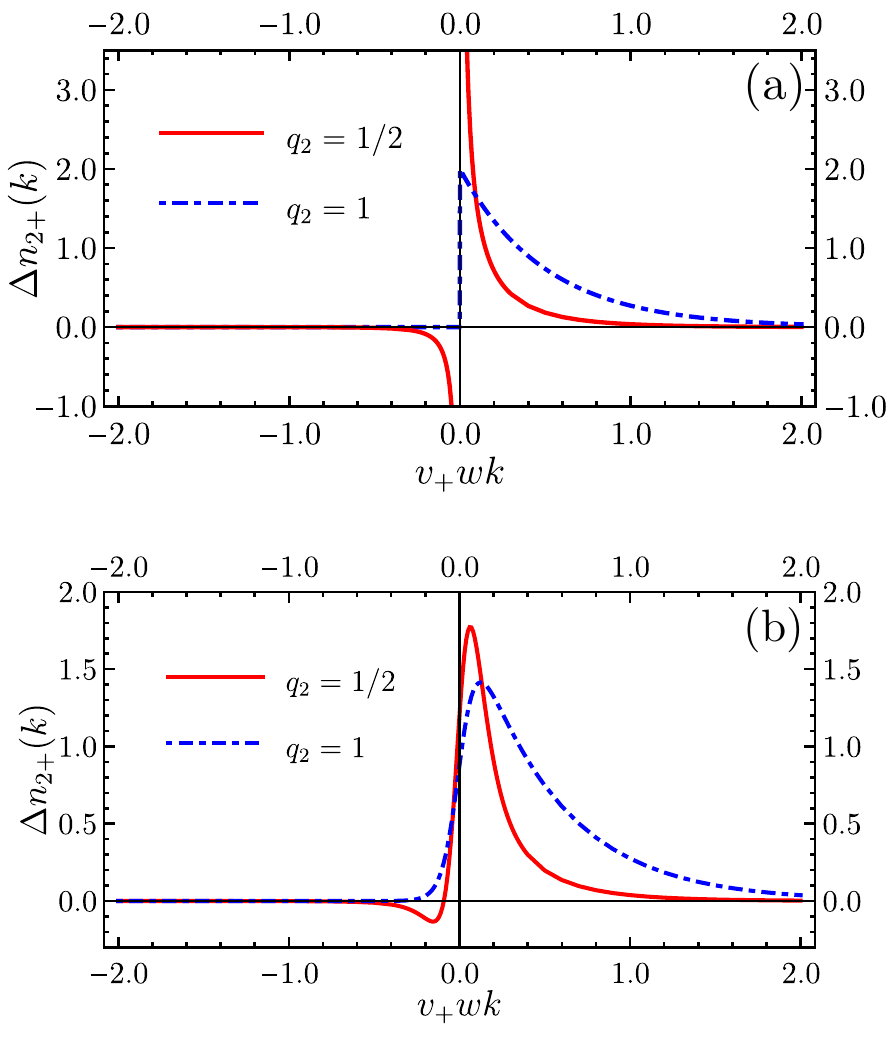}
		\caption{Distribution $\Delta n_{2+}(k)$ (in units of $wv_+$) at fixed interaction angle parameter $\theta=\pi/4$. In this case the charge carried by the fractional excitation is $-eq_2=-eq/2$, which is an integer multiple of $-e$ for $q\in 2\mathbb{N}$. (a) Zero temperature result; for integer charges, the distribution is given by \eqref{eq:nk-stationary} and describes a pure electronic excitation, with no associated holes. In contrast, when $q_2$ is not an integer, a divergence appears near the Fermi momentum. (b) Finite temperature distributions for $Tw=0.05$ ($T$ is the temperature). Note that the divergence near the Fermi momentum is washed out. Still, hole contributions do appear in the case of non integer charges, while in the integer case the distribution is always positive.}
		\label{fig:nk}
	\end{center}
\end{figure}
It is easy to verify that the total charge of each wave packet [see Eq. \eqref{eq:frac-charges}] is recovered by integrating its distribution over momentum $k$. The distribution of a given excitation has thus an exponential profile, modulated by a polynomial. More importantly, previous expressions indicate that both wave packets on the outer channel and the fast one on the inner channel are made only of electron excitations above the Fermi level, while the slow wave packet on the inner channel is made only of hole excitations below the Fermi level. It is worth underlining that this feature is peculiar of Lorentzian voltage pulses and it is not shared by generic wave packets generated with other drives, which would contain electron-hole pair contributions.
As a last comment, we note that the charges of the outer-channel excitations can be simultaneously integer only if $q$ itself is integer, since $q_{1+}=m_+$ and $q_{1-}=m_-$ imply $q=m_++m_-$. This condition, however, can be achieved only for particular values of the mixing angle, such that $\tan^2\!\theta=m_+/m_-$, in agreement with Ref.\ \onlinecite{grenier13}. On the inner channel, instead, given any interaction strength it is possible to have both fractionalized excitations with integer charge if $q=2n/\sin2\theta$, with $n\in\mathbb{N}$ and without further constraints. This means that, by finding a quantity showing clear signatures whenever the charge of fractionalized excitations is integer, and knowing the values of $q$ for which this happens, the mixing angle can be extracted from the last relation. This will thus provide information on the strength of interactions in copropagating edge channels. This idea will be further developed in Sec.\ \ref{sec:noise-per}.

At this point, we conclude the discussion of the stationary regime with the general case where the charge of a given excitation is not an integer multiple of $-e$. In this situation the momentum distribution features particle-hole pair contributions. This is shown in Fig.\ \ref{fig:nk}(a), where we plot the zero-temperature distribution $\Delta n_{2+}(k)$ for $\theta=\pi/4$ and compare the case of integer $(q_2=1)$ and non-integer $(q_2=1/2)$ charges. In the latter case the distribution is evaluated numerically and clearly features a divergence around $k=0$, with negative contributions for $k<0$, thus signaling the presence of hole excitations. We do not show $\Delta n_{2-}$ since its behavior is analogous to the one just described, with electron contributions replaced by holes and vice versa. Indeed the two distributions are linked by $v_+\Delta n_{2-}(k/v_-)=-v_-\Delta n_{2+}(-k/v_+)$

The divergence around $k=0$ is common to all distributions $\Delta n_{\alpha\pm}$, whose scaling behavior is analytically obtained by asymptotically evaluating them for small $k$, i.e. close to Fermi momentum. We have (see Appendix \ref{app:scaling} for details)
\begin{equation}
\Delta n_{\alpha\pm}(k\to 0)\approx\frac{1-\cos(2\pi q_{\alpha\pm})}{2\pi^2}\frac{1}{k}\,,
\label{eq:scaling}
\end{equation}
with $q_{\alpha\pm}$ defined in \eqref{eq:frac-charges}.
This is precisely the particle-hole pair contribution appearing for non-integer $q_{\alpha\pm}$. We see that the number of particle-hole pairs exhibit a logarithmic divergence\cite{levitov96,levitov97,keeling06}, as it is clear by integrating the distributions over $k$. This is a manifestation of the orthogonality catastrophe for fermions\cite{levitov96}. Finally, in Fig.\ \ref{fig:nk}(b) we present finite-temperature results, showing that the zero-temperature divergence disappears and the distributions are smeared around $k=0$.

\subsection{Transient regime}
Although the non-equilibrium momentum distribution in the stationary regime described so far can give important information on its own, interesting effects can arise also when considering the finite time regime with partially overlapping pulses. We shall investigate this regime by focusing on the inner channel, since it is more sensitive in order to probe interactions effect with respect to the outer one. 

When slow and fast wave packets are not well separated, the full correlator \eqref{eq:correlator-et-2} must be used to compute the momentum distribution $\Delta n_2$, without resorting to the separation \eqref{eq:separation}. However, it is always possible to write the whole momentum distribution of the inner channel as
\begin{equation}
\Delta n_2(k,t)=\Delta n_{2+}(k)+\Delta n_{2-}(k)+\Delta n_{2X}(k,t)\,,
\end{equation}
where $\Delta n_{2\pm}(k)$ are the stationary distributions of fractional excitations considered independently, as discussed in Sec.\ \ref{sec:nk-stationary}, while $\Delta n_{2X}$ is a ``crossed'' term carrying all information on their overlap. Note that the time dependence is carried by the overlap term.
\begin{figure}[t]
	\begin{center}
		\includegraphics[width=\columnwidth]{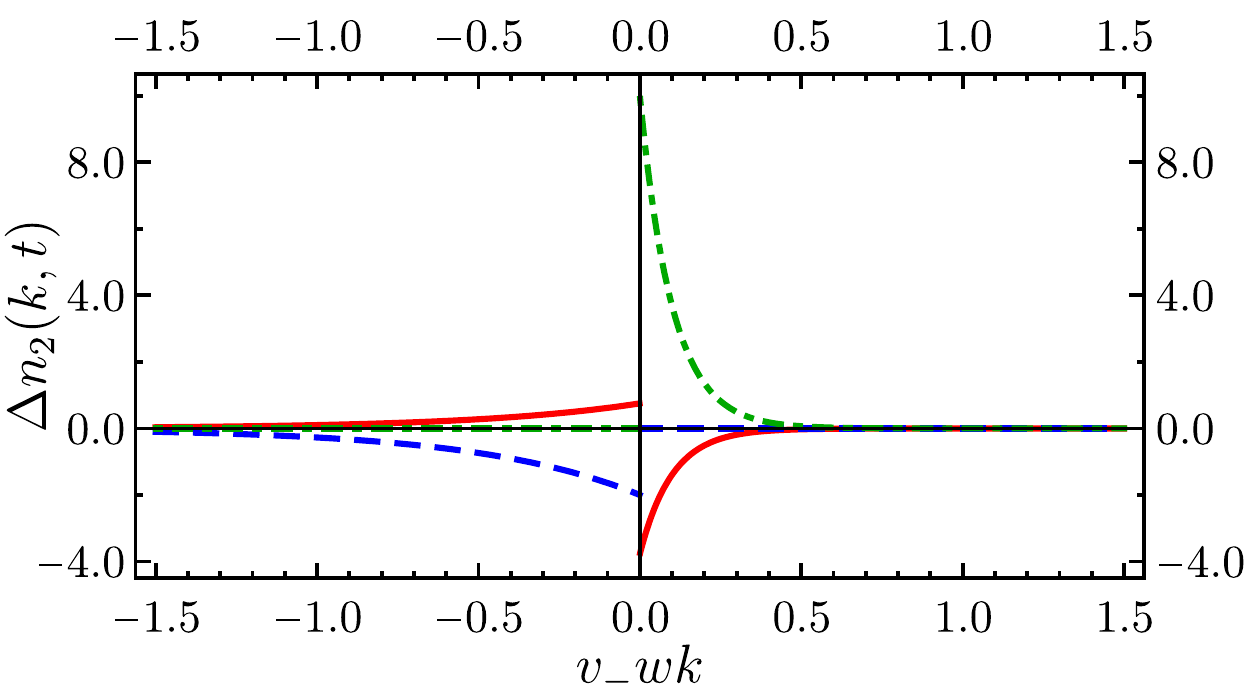}
		\caption{Overlap distribution \eqref{eq:nkx-1}, together with stationary ones \eqref{eq:nk-stationary}, in the case $q_2=1$. The blue (green) dashed (dot-dashed) line represents $\Delta n_{2-}$ $(\Delta n_{2+})$, while the red full line is $20\Delta n_{2X}$ for $t/w=8$. Note that the latter describes electron (hole) excitations for negative (positive) momenta, in contrast with the behavior of the stationary distributions \eqref{eq:nk-stationary}. In this plot we have used $\gamma=1/5$.}
		\label{fig:nkx}
	\end{center}
\end{figure}
It reads
\begin{equation}
\begin{aligned}
\Delta n_{2X}(k,t)&=\frac{1}{2\pi}\int_{-\infty}^{+\infty}dx\int_{-\infty}^{+\infty}d\xi\,e^{-ik\xi}\\
&\times\left[\Delta\mathscr{G}_2(x,\xi;t)-\sum_{\eta=\pm}\Delta\mathscr{G}_{2\eta}(x_\eta,\xi)\right]\,,
\end{aligned}
\label{eq:nkx-gen}
\end{equation}
where the correlators appearing in the above equation are given in \eqref{eq:correlator-et-2} and \eqref{eq:separated-correlators}, and in general it has to be evaluated numerically. To better appreciate the overlap contribution, it is instructive to specialize to the case of a Lorentzian pulse with ${q_2}=n\in\mathbb{N}$. In this case, we find the following expression in terms of an integral over energies $\omega$ (see Appendix \ref{app:nk-int})
\begin{equation}
\begin{aligned}
&\Delta n_{2X}(k,t)=\\
&\gamma\sum_{r,p=1}^{n}\partial_k\left|\int_{-\infty}^{+\infty}\frac{d\omega}{2\pi}\,\tilde{\chi}_r(\omega)\tilde{\chi}_p^*(\gamma\omega-v_-k)e^{-i\omega t(1-\gamma)}\right|^2\,,
\end{aligned}
\label{eq:nk-crossed}
\end{equation}
where we have introduced the ratio $\gamma\equiv v_-/v_+<1$ and the functions
\begin{equation}
\tilde{\chi}_p(\omega)=2i\sqrt{w\pi}\Theta(\omega)L_{p-1}(2w\omega)e^{-w\omega}\,.
\label{eq:tilde-chi}
\end{equation}
As a simple example we consider the case ${q_2}=1$, where the result reads
\begin{equation}
\Delta n_{2X}(k,t)=-\frac{8w\gamma (\gamma+1)^{-2}}{1+\frac{t^2}{w^2}\left(\frac{1-\gamma}{1+\gamma}\right)^2}
\sum_{\eta=\pm}\eta v_\eta\Theta(\eta k)e^{-2\eta v_\eta wk}\,.
\label{eq:nkx-1}
\end{equation}
As shown in Fig.\ \ref{fig:nkx}, this mixed term describes electron excitations $(\Delta n_{2X}>0)$ at $k<0$ and hole excitations $(\Delta n_{2X}<0)$ at $k>0$. This means that the effect of the overlap between two oppositely-charged pulses carrying integer charges results in an effective reduction of the overall number of electron and hole excitations, with respect to the case of completely separated wave packets. Note that the time-dependent overlap contribution $\Delta n_{2X}(k,t)$ becomes negligible at times $t\gg w\frac{1+\gamma}{1-\gamma}$, which is precisely the relation introduced at the beginning of Sec.\ \ref{sec:nk-stationary} as the condition allowing us to consider fractional excitations well separated and independent.

\begin{figure}[t]
	\begin{center}
		\includegraphics[width=\columnwidth]{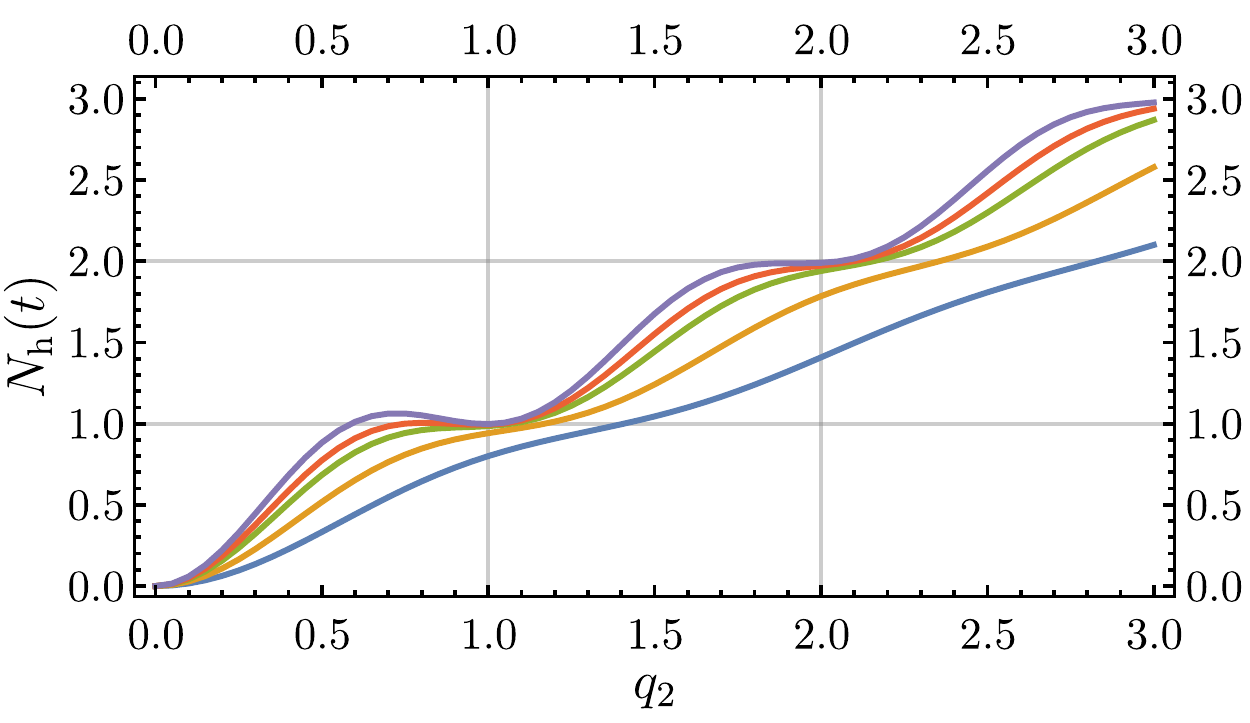}
		\caption{Total number of injected holes as a function of ${q_2=q\cos\theta\sin\theta}$ at zero temperature, in the case of a Lorentzian drive. Different curves refer to (from bottom to top) $t/w=5,10,20,30,50$. The ratio between slow and fast velocities is $\gamma=1/5$.}
		\label{fig:nh-sp}
	\end{center}
\end{figure}

In order to better characterize the particle-hole pair production due to the overlap contribution in the transient regime, we compute the number of holes $N_\text{h}(t)$ generated at time $t$ as a consequence of the applied voltage. Since hole contributions are given by negative values of the momentum distribution, $N_\text{h}(t)$ is 
\begin{equation}
N_\text{h}(t)=-\int_{-\infty}^{+\infty}dk\,\Delta n_2(k,t)\Theta[-\Delta n_2(k,t)]\,.
\end{equation}
At zero temperature, the previous expression reduces to an integration over negative momenta. In this situation we obtain (see Appendix \ref{app:nh} for details)
\begin{equation}
	\begin{aligned}
	&N_\text{h}(t)=\frac{1}{(2\pi)^2}\int_{-\infty}^{+\infty}dx\int_{-\infty}^{+\infty}dy\frac{1}{(a+iy)^2}\\
	&\times\cos\left[e\frac{q_2}{q}\left(\int_{t_++\frac{y}{2v_+}}^{t_+-\frac{y}{2v_+}}dt'V(t')-\int_{t_-+\frac{y}{2v_-}}^{t_--\frac{y}{2v_-}}dt'V(t')\right)\right]\,,
	\label{eq:nh-T0}
	\end{aligned}
\end{equation}
with $t_\pm=t-x/v_\pm$.
The behavior of $N_\text{h}(t)$ as a function of ${q_2}$ and for different times is reported in Fig.\ \ref{fig:nh-sp}, where we again consider the Lorentzian drive \eqref{eq:lor-pulse}. Note that, upon increasing $t/w$, $N_\text{h}(t)\to q_2$ when $q_2$ is integer, as expected from the previous analysis of completely separated pulses. For intermediate times, instead, this quantity reflects the effective charge reduction discussed above due to the overlap contribution $\Delta n_{2 X}(k,t)$.
Moreover, for non integer values of ${q_2}$, we clearly see that $N_\text{h}$ grows upon increasing $t/w$. For sufficiently large $t/w$, this increase is logarithmic in $t$ or, equivalently, in the propagation distance along the edge. This feature agrees with the scaling behavior \eqref{eq:scaling}, appearing in the stationary regime: Eq. \eqref{eq:scaling} indeed results in a logarithmic divergence of the number of produced holes.

\section{Noise and particle-hole pair production}\label{sec:noise}
Particle-hole pair production discussed above can be experimentally accessed through low-frequency noise measurements. It is indeed in this way that Levitons were proved to be minimal excitations in non-interacting quantum conductors\cite{keeling06,dubois2013levitonsNature}.
We will here compute the noise generated when fractional excitations are partitioned, after a propagation distance $d$, by a QPC allowing tunneling of electrons between the two edges of the Hall bar, as shown in Fig.\ \ref{fig:setup}. This is known as the Hanbury-Brown Twiss (HBT) configuration. The noise generated by partitioning outer-channel excitations at a QPC has been considered in some works in the literature\cite{grenier13,milletari13,inoue2014}. Here, we focus on what happens in the inner channel, since it is more suitable to probe interactions, as we will see in the following. To this end, the QPC is assumed to be polarized so as to completely transmit the outer channel and weakly reflect the inner one. We shall therefore use a tunneling Hamiltonian
\begin{equation}
H_\text{T}=\Lambda\Psi_{R2}^\dagger(0)\Psi_{L2}(0)+\text{H.c.}\,,
\end{equation}
to be considered as a perturbation, $\Lambda$ being a small constant amplitude. The quantity we are interested in is the noise measured at terminal 2, see Fig.\ \ref{fig:setup}(a). At first we shall consider the noise produced by a single voltage pulse; then we will turn our attention to the experimentally more relevant situation of a periodic train of pulses.

\subsection{Single voltage pulse}\label{sec:noise-sp}
We define the zero frequency noise as\cite{dubois13prb,grenier13,ferraro2014noise,moskalets17,ferraro10tunneling}
\begin{equation}
\begin{aligned}
S=&2\int_{-\infty}^{+\infty}\!\!\!\!dt\int_{-\infty}^{+\infty}\!\!\!\!d\tau\left[\Braket{J(t+\tau)J(t)}-\Braket{J(t+\tau)}\Braket{J(t)}\right]\,,
\end{aligned}
\label{eq:definition-noise}
\end{equation}
where $J(t)=-J_{L2}(-d,t)$, with $J_{L2}$ the current operator of the inner channel on the left-moving edge [see Fig.\ \ref{fig:setup}(a)]. Note that the sign has been chosen in such a way that the current flowing into terminal 2 is taken as positive. The time evolution of current operators will be computed at lowest order in the tunneling. Another useful quantity is the excess noise $\Delta S=S-2e\overline{\Braket{J(t)}}$, with $\overline{\Braket{J(t)}}=\int_{-\infty}^{+\infty}dt\Braket{J(t)}$, physically measuring deviations from the Poissonian value. However, in the particular case we are considering, $\overline{\Braket{J(t)}}=0$ since it represents the total charge flowing to the terminal 2 and this quantity vanishes because the current on the inner channel is made of two pulses carrying opposite charges.
 We can therefore refer to noise or excess noise interchangeably. The calculation of the excess noise, showing that $\Delta S=S$, is provided in Appendix \ref{app:noise}. Here we simply state the result:
\begin{equation}
	\begin{aligned}
	S&=\frac{4e^2|\Lambda|^2}{(2\pi a)^2}\int_{-\infty}^{+\infty}dt\int_{-\infty}^{+\infty}d\tau\left(\frac{a}{a+iv_-\tau}\frac{\pi T\tau}{\sinh\pi T\tau}\right)^2\\
	&\times\cos\left[e\frac{q_2}{q}\left(\int_{t-\tau}^{t}dt'V(t')-\int_{t-\tau+\tau_d}^{t+\tau_d}dt'V(t')\right)\right]\,,
	\label{eq:noise-sp}
	\end{aligned}
\end{equation}
where $\tau_d=d(v_+^{-1}-v_-^{-1})$.

\begin{figure}[t]
	\begin{center}
		\includegraphics[width=\columnwidth]{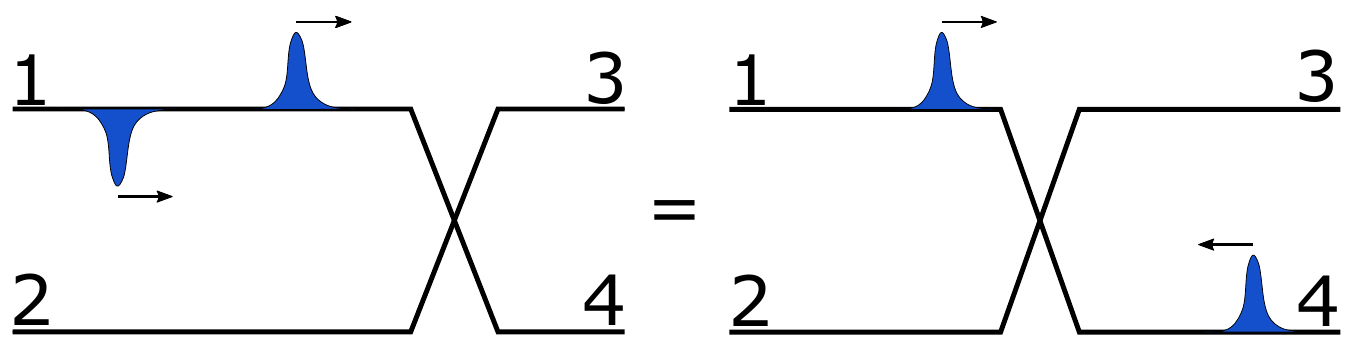}
		\caption{Equivalence between the investigated HBT setup at $\nu=2$ (left) and a non-interacting HOM setup at $\nu=1$ (right). In the first one, the two oppositely-charged excitations on the inner channel arrive at a QPC from the left side and are separated due to the fractionalization phenomena. In the second one, two excitations with the same charge arrive at a QPC from opposite sides and are separated due to the time delay between the drive applied at terminal 1 and the one at terminal 4.}
		\label{fig:hom-hbt}
	\end{center}
\end{figure}

There are a few noteworthy aspects in the above formula. First of all, Eq.\ \eqref{eq:noise-sp} describes the noise generated when two identical but oppositely charged excitations arrive at the same side of the QPC, separated by a time $\tau_d$, due to the fractionalization mechanism. This result can be formally mapped onto the one obtained with a HOM setup in a \emph{non-interacting system}. We recall that in a HOM configuration, both terminals 1 and 4 [refer to Fig.\ \ref{fig:setup}(a)] are driven by external voltages, with a tunable time delay between the drive applied at terminal 1 and the one at terminal 4. With this setup, Eq.\ \eqref{eq:noise-sp} can be obtained in a non-interacting system where two identical excitations, generated from terminals 1 and 4 and carrying \emph{the same} charge $-e{q_2}$, arrive at the QPC from opposite sides and with a time delay $\tau_d$. This equivalence is sketched in Fig.\ \ref{fig:hom-hbt}. Our HBT setup thus simulates a HOM interferometry of fractional excitations, where the time delay is controlled by the propagation distance along the edge and the different propagation velocities.

Secondly, the zero temperature limit of Eq.\ \eqref{eq:noise-sp} resembles the expression \eqref{eq:nh-T0} for  $N_\text{h}$, even though they are not exactly the same. However, the analogy between the two expressions suggests that, upon specifying a particular voltage, a relation between the time $t$ appearing in $N_\text{h}$ and the time delay $\tau_d$ in $S$ can be found so that noise and number of holes are proportional. 
For the Lorentzian drive \eqref{eq:lor-pulse}, this is indeed the case: for
\begin{equation}
\frac{t^\star}{w}=\sqrt{\frac{\gamma}{(\gamma-1)^2}\left(\frac{\tau_d}{w}\right)^2-1}
\label{eq:link-nh-noise}
\end{equation}
we find
\begin{equation}
\frac{S}{S_0}=2N_\text{h}(t^\star)\,,
\end{equation}
where $S_0=2e^2|\Lambda|^2/v_-^2$ is a reference noise value. Note that $|\Lambda|^2/v_-^2$ can be interpreted as the reflection probability of the QPC.
The noise as a function of ${q_2}$ presents the same behavior as shown in Fig.\ \ref{fig:nh-sp}. 
In particular, $S/S_0$ approaches the value $2{q_2}$ at integer ${q_2}$, upon increasing the propagation distance $d$. This is very clearly understood from the HOM perspective: the greater the propagation distance, the better the two wave packets are separated and the greater is $\tau_d$. From the HOM point of view, this means that two Levitons arrive at the QPC separated by a very long time and therefore they contribute independently to the noise, which becomes twice as big as the one a single Leviton would produce. Instead, when the overlap between the two wave packets is not negligible, the noise is reduced because of the anti-bunching effect\cite{dubois13prb}. This is fully consistent with what we have found in the previous section discussing the behavior of the nonequilibrium momentum distribution in the transient regime. 
Finally, we recall that for other drives than the Lorentzian pulse, the noise produced by two well separated excitations carrying integer charges would be bigger than $2{q_2}$, since each excitation is accompanied by a cloud of particle-hole pairs contributing to it.

\subsection{Periodic train of pulses}\label{sec:noise-per}
Electron quantum optics experiments rely on periodic voltage drives, instead of single shot measurements\cite{bocquillon14,dubois13prb}. To make more connection with possible experimental implementation, we thus turn our attention precisely to this situation: from now on, the voltage drive will be a periodic train of pulses, so that $V(t)=V(t+\mathcal{T})$. Therefore, we have to properly modify the definition \eqref{eq:definition-noise}. This can be done by taking the average of the noise over one period of the drive, which formally amounts to replace $\int_{-\infty}^{+\infty}dt\,$ with $\int_{0}^{\mathcal{T}}\frac{dt}{\mathcal{T}}$. In the same way, $\overline{\Braket{J(t)}}$ becomes the time-average of the current over the period of the drive. Moreover, in the definition \eqref{eq:definition-q} of the parameter $q$, we substitute $\int_{-\infty}^{+\infty}dt\,\to\int_{0}^{\mathcal{T}}dt$ so that $-eq$ becomes the charge \emph{per period} carried by the pulse $V(t)$. It can be linked to the dc component of the voltage, namely $V_\text{dc}=\int_0^\mathcal{T}\frac{dt}{\mathcal{T}}V(t)$, by the relation $q\Omega=-eV_\text{dc}$, with $\Omega=2\pi\mathcal{T}^{-1}$ the angular frequency of the drive. Thanks to the periodicity of the drive, the phase factor $\exp[ ie\int_{-\infty}^{t}dt'V_\text{ac}(t')]$, with $V_\text{ac}(t)=V(t)-V_\text{dc}$ the ac component of the drive, is also periodic and can be thus expanded in a Fourier series:
\begin{equation}
e^{ie\int_{-\infty}^{t}dt'V_\text{ac}(t')}=\sum_{\ell=-\infty}^{+\infty}p_\ell\, e^{-i\ell\Omega t}\,.
\label{eq:pl}
\end{equation}
Coefficients $p_\ell$ are called photo-assisted amplitudes\cite{dubois13prb}; for $\ell<0$ $(\ell>0)$ they give the probability amplitude for an electron to emit (absorb) $|\ell|$ photons. They satisfy $\sum_{\ell}|p_\ell|^2=1$ and $\sum_{\ell}\ell\,|p_\ell|^2=0$. By using this formalism, the noise \eqref{eq:noise-sp} (after modifying the first time integral into a $\mathcal{T}$-average) can be written in the following form
\begin{equation}
\frac{S}{S_0\Omega}=\frac{1}{2\pi}\sum_{\ell=-\infty}^{+\infty}|\tilde{p}_\ell({q_2},\tau_d)|^2\,\ell\,\coth\left(\frac{\ell\Omega}{2T}\right)\,,
\end{equation}
where coefficients $\tilde{p}_\ell$ read
\begin{equation}
\tilde{p}_\ell({q_2},\tau_d)=\sum_{m=-\infty}^{+\infty}p_{\ell+m}({q_2})p_\ell^*({q_2})e^{-i2\pi\tau_d/\mathcal{T}}\,.
\end{equation}
As already mentioned, these coefficients are precisely the ones we would obtain in a non-interacting HOM setup where two excitations carrying the same charge $-e{q_2}$ collide at the QPC, with a delay $\tau_d/\mathcal{T}$. The main difference with respect to the single pulse voltage discussed in Sec.\ \ref{sec:noise-sp} is that now the result is periodic in $\tau_d/\mathcal{T}$ and the maximal separation between two consecutive pulses is achieved for $\tau_d=\mathcal{T}/2$. The actual parameter controlling the separation between the wave packets in the periodic case is the ratio $\eta$ between the temporal extension of each pulse of the drive and its period: $\eta=w/\mathcal{T}$. In the following we will consider two different drives.

As a first example, we choose a periodic train of Lorentzian pulses
\begin{equation}
V(t)=\frac{V_\text{dc}}{\pi}\sum_{p=-\infty}^{+\infty}\frac{\eta}{\eta^2+(t/\mathcal{T}-p)^2}\,,
\label{eq:lor-periodic}
\end{equation}
\begin{figure}[tbp]
	\begin{center}
		\includegraphics[width=\columnwidth]{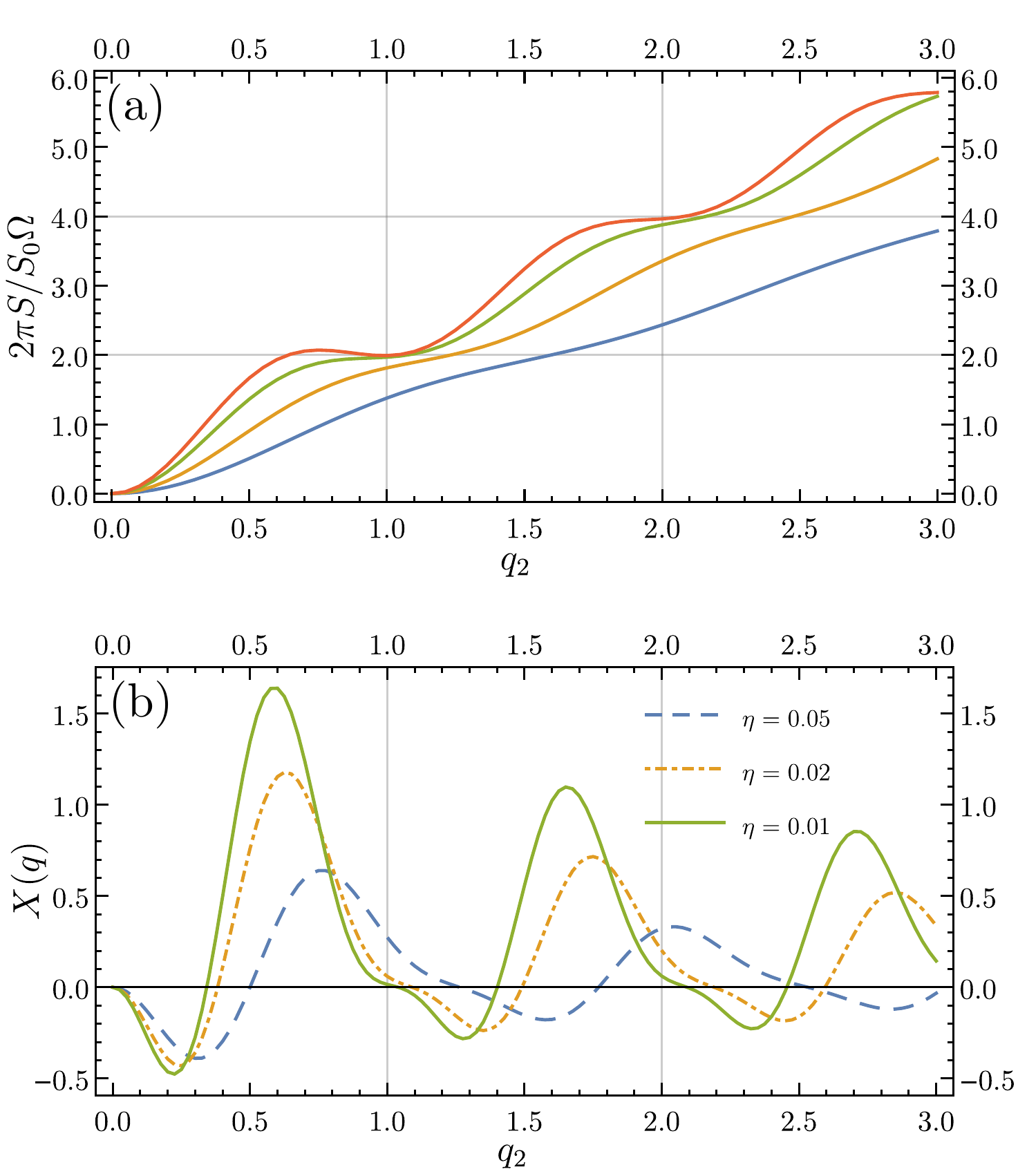}
		\caption{Panel (a): inner-channel noise $2\pi S/(S_0\Omega)$ for the periodic Lorentzian drive \eqref{eq:lor-periodic}, as a function of ${q_2=q\cos\theta\sin\theta}$, for $\tau_d=\mathcal{T}/2$ and (from bottom to top) $\eta=0.1,\,0.05,\,0.02,\,0.01$. The value $2{q_2}$ is reached at integer ${q_2}$ upon lowering $\eta$. Panel (b): the function \eqref{eq:function-x} at zero temperature and $\tau_d=\mathcal{T}/2$ for the Lorentzian drive \eqref{eq:lor-periodic}. We note that, upon decreasing $\eta$, the zeros (in the neighborhood of which $X$ is decreasing) are more and more precisely localized at integer values of $q_2$.}
		\label{fig:noise-periodic-lor}
	\end{center}
\end{figure}
whose photo-assisted amplitudes are\cite{dubois13prb,rech16prl}
\begin{equation}
p_\ell(q)=qe^{-2\pi\eta\ell}\sum_{s=0}^{+\infty}\frac{\Gamma(q+\ell+s)}{\Gamma(q+1-s)}\frac{(-1)^se^{-4\pi\eta s}}{s!(\ell+s)!}\,.
\end{equation}
We show in Fig.\ \ref{fig:noise-periodic-lor}(a) the zero-temperature noise for $\tau_d=\mathcal{T}/2$ and different values of $\eta$. We observe a behavior similar to the single-pulse noise reported in Fig.\ \ref{fig:nh-sp}, though quantitatively different. In particular, for $\eta\to 0$, we recover that the noise reaches the value $2{q_2}$ at integer ${q_2}$. Here we would like to stress another feature: we note that, upon decreasing $\eta$, $S$ has an increasingly well defined staircase behavior, the period of the steps being ${q_2}$.
This important observation can be exploited to construct a measurable quantity giving direct access to the mixing angle $\theta$. Let us explain how this idea can be developed. We introduce the function
\begin{equation}
X({q})=2\pi\,\frac{2S({q})-S(2{q})}{S_0\Omega}\,.
\label{eq:function-x}
\end{equation}
In the limit $\eta\to 0$, $X({q_2}\in\mathbb{N})=0$ for the Lorentzian drive \eqref{eq:lor-periodic}, indicating that it is a natural candidate to be exploited.
Indeed, one can realize that by plotting $X$ as a function of the tunable parameter $q$ and looking at the position of its zeros, the mixing angle $\theta$ could be directly extracted. In Fig.\ \ref{fig:noise-periodic-lor}(b) we show the function \eqref{eq:function-x} at zero temperature for a Lorentzian drive with different values of $\eta$.
It presents two different classes of zeros, depending whether the function is increasing or decreasing. We refer to those in the neighborhood of which the function decreases. While it is clearly visible that upon decreasing $\eta$, the zeros are more and more precisely localized at integer values of $q_2$, we note that significant deviations from this ideal situation already appear for $\eta=0.05$. These deviations are further enhanced by displacing $\tau_d$ from the optimal value $\mathcal{T}/2$ as well as by finite-temperature effects. Therefore, from a practical point of view, the Lorentzian drive cannot be used for the purpose of extracting the mixing angle $\theta$.

\begin{figure}[tbp]
	\begin{center}
		\includegraphics[width=\columnwidth]{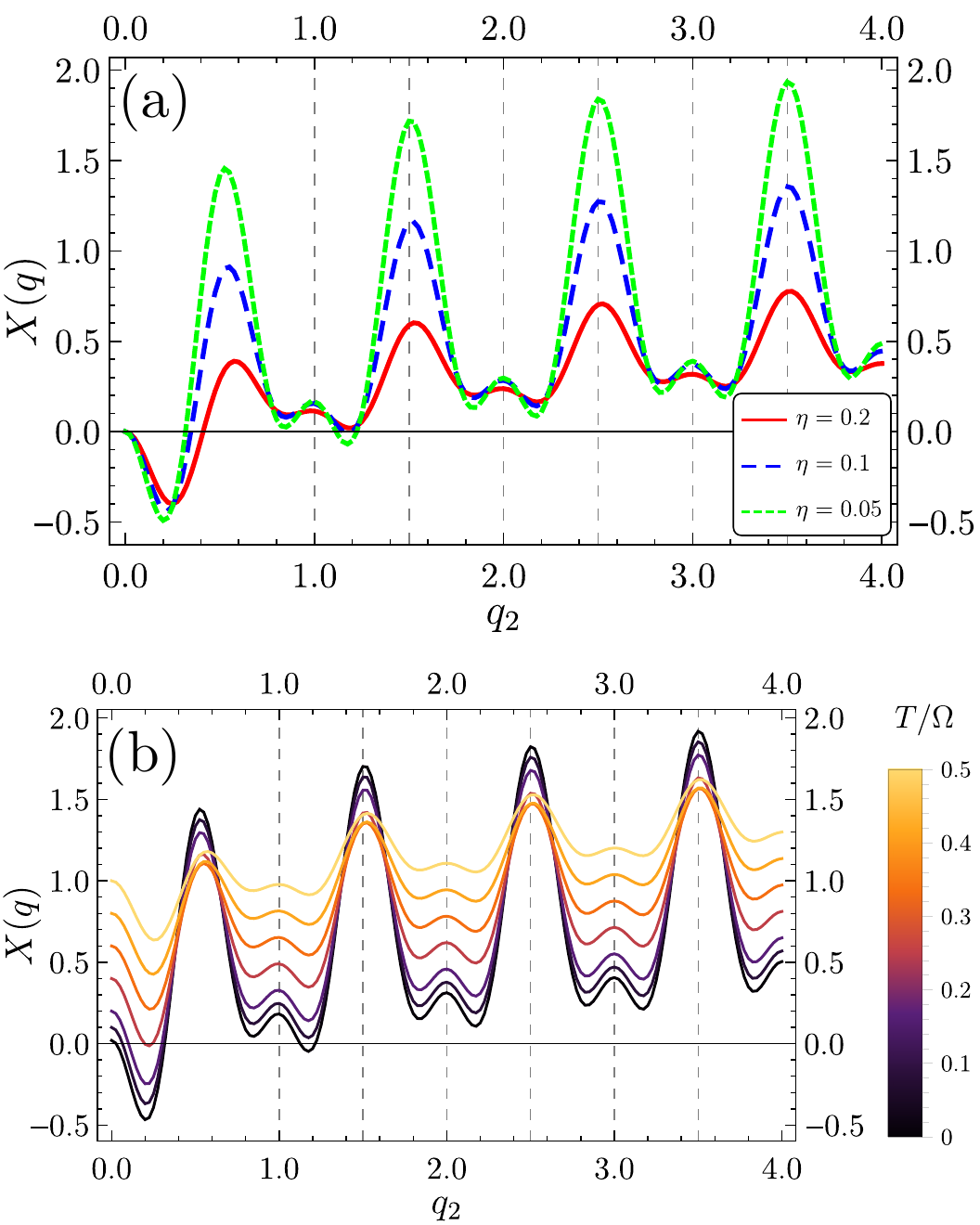}
		\caption{The quantity $X$, defined in \eqref{eq:function-x}, as a function of ${q_2}$ for a rectangular pulse. (a) Zero-temperature result for different values of $\eta$. Principal maxima are very well located at half-integer ${q_2}$ already for $\eta=0.1$. In addition, the smaller $\eta$ the bigger the amplitude of oscillations. (b) Different curves for temperatures ranging from zero to $T=\Omega/2$ and $\eta=0.05$. In both panels the time delay is $\tau_d=\mathcal{T}/2$.} 
		\label{fig:noise-rect}
	\end{center}
\end{figure}

However, one may wonder if other drives, though not generating clean pulses for integer values of $q_2$, still exhibit some signatures in the function \eqref{eq:function-x} at those values, allowing us to directly probe the value of $\theta$. The answer is affirmative: clear and stable signatures are found for a train of rectangular pulses. For each period, the signal is
\begin{equation}
V(t)=\frac{V_\text{dc}}{2\eta}[\Theta(t)+\Theta(\eta\mathcal{T}-t)]\,,\quad t\in[0,\mathcal{T})\,.
\end{equation}
Here, $\eta$ represents the width of the rectangular pulse in units of the period. The photo-assisted coefficients for this drive are given by\cite{ferraro18squeezing}
\begin{equation}
p_\ell(q)=\frac{q}{\pi}\frac{e^{i\pi[\eta\ell+q(\eta-1)]}\sin\{\pi[\eta\ell+q(\eta-1)]\}}{(q+\ell)[\eta\ell+q(\eta-1)]}\,.
\end{equation}
In Fig.\ \ref{fig:noise-rect}(a) we show $X({q})$ for this drive, in the case $\tau_d=\mathcal{T}/2$ and at zero temperature. We observe, after a transient at small values of ${q_2}$, a regular oscillating pattern, with local maxima in correspondence of integer ${q_2}$ and principal maxima better and better located at half-integer values of ${q_2}$ the more $\eta$ is decreased.
When $\tau_d$ differs from the optimal value $\mathcal{T}/2$ principal maxima are less precisely localized at half-integer ${q_2}$, this effect being more and more irrelevant the smaller $\eta$ is. For instance, if $\eta=0.1$ and $\tau_d=\mathcal{T}/4$, we simply have a reduction of the amplitude of oscillation, while the position of both principal and secondary maxima is substantially unaffected (not shown). Finally, finite temperature progressively reduces the amplitude of oscillations, but has very little influence on the position of the maxima, as it can be seen in Fig.\ \ref{fig:noise-rect}(b).

All features discussed above enable to determine the value of the mixing angle $\theta$ from the quantity $X$ introduced in \eqref{eq:function-x}, by plotting it as a function of the tunable external parameter $q$ and looking for the values of $q$ at which $X$ has principal or secondary maxima. For those values the charge number of fractional excitations, i.e.\ $q_2=q\cos\theta\sin\theta$, must be integer or half-integer, thus allowing to extract the parameter $\theta$.

\section{Conclusions}\label{sec:conclusions}
In this work we have considered the nonequilibrium dynamics of few-electron wave packets in the ballistic edge channels of the integer quantum Hall effect at filling factor $\nu=2$ in presence of time-dependent external drives.
Screened Coulomb interactions between copropagating channels strongly affect the dynamics, leading to the fractionalization of wave packets into collective excitations propagating at different velocities along the edges. In particular, on the inner channel two oppositely charged excitations emerge. We have analyzed in detail their out-of-equilibrium momentum distributions, showing that for properly designed Lorentzian pulses, they vanish for momenta smaller (greater) than the Fermi momentum when the number of electron charges carried by each fractional excitation is a positive (negative) integer. This is in contrast with the case of non-integer charges where a cloud of particle-hole pairs is produced. They are distributed as $1/k$ close to the Fermi momentum, leading to a logarithmic divergence of the number of excitations produced by the external pulse, which we have inspected by computing the number of holes $N_\text{h}$ directly from the momentum distribution. By focusing on the inner-channel excitations, we have investigated effects linked to their spatial overlap, interpreting the results from the point of view of an equivalent HOM interferometry between fractional excitations.
Finally, by considering a periodic drive, we have showed that rectangular pulses constitute a powerful tool, suitable to extract the mixing angle $\theta$ controlling the strength of interactions between edge channels. This makes our results relevant for the investigation of correlations and the microscopic nature of interactions in quantum Hall edge channels.

\begin{acknowledgements}
	M.A.\ and M.S.\ acknowledge support from the University of Genova through grant No.\ 100020-2017-MS-FRA\_001.
	This work was granted access to the HPC
	resources of Aix-Marseille Universit\'e financed by the
	project Equip@Meso (Grant No.\ ANR-10-EQPX29-01). It has been carried out in the framework of project ``one shot reloaded'' (Grant No.\ ANR-14-CE32-
	0017) and benefited from the support of the Labex ARCHIMEDE (Grant No.\ ANR11-LABX-0033) and the AMIDEX project (Grant No.\ ANR-11-IDEX-0001-02), funded by the ``investissements d'avenir'' French Government program managed by the French National Research Agency (ANR).
	M.C.\ acknowledges support from the Quant-EraNet project Supertop and the CNR-CONICET cooperation programme ``Energy conversion in quantum, nanoscale, hybrid devices''.
 \end{acknowledgements}
\appendix
\section{Calculation of momentum distributions for integer charges}\label{app:nk-int}
In this Appendix we derive formulas \eqref{eq:nk-stationary} and \eqref{eq:nk-crossed}. Let us begin with the first one (we prove the formula for $\Delta n_{2\pm}$; the proof for $\Delta n_{1\pm}$ is identical). The starting point is Eq.\ \eqref{eq:corr-lor}, which we rewrite here (recall that this is a zero-temperature expression)
\begin{equation}
\Delta\mathscr{G}_{2\pm}=\frac{1}{2\pi i \xi}\left[\prod_{\eta=\pm}\left[\frac{iv_\pm w-\eta(v_\pm t_\pm+\eta\xi/2)}{iv_\pm w+\eta(v_\pm t_\pm+\eta\xi/2)}\right]^{\pm{q_2}}-1\right]\,,
\end{equation}
where ${q_2}=q\cos\theta\sin\theta$ and $t_\pm=t-x/v_\pm$. When ${q_2}=n\in\mathbb{N}$, the following representation holds\cite{moskalets15,glattli16wavepackets,ronetti17levitons}:
\begin{equation}
\Delta\mathscr{G}_{2\pm}=\frac{\pm 1}{v_\pm}\sum_{j=1}^{n}\chi_j^\pm\left(t_\pm-\frac{\xi}{2v_\pm}\right)\chi_j^\mp\left(t_\pm+\frac{\xi}{2v_\pm}\right)\,,
\label{eq:app:chi1}
\end{equation}
where $\chi_j^+\equiv\chi_j$, $\chi_j^-\equiv\chi_j^*$ and
\begin{equation}
\chi_j(t)=\sqrt{\frac{w}{\pi}}\frac{(t+iw)^{j-1}}{(t-iw)^j}\,.
\end{equation}
Eq.\ \eqref{eq:app:chi1} can be checked by direct calculation for $n=1$ and then proved in general by induction. Now we substitute \eqref{eq:app:chi1} into \eqref{eq:nk}, introduce the Fourier transform
\begin{equation}
\begin{aligned}
\tilde\chi_j(\omega)&=\int_{-\infty}^{+\infty}dt\,\chi_j(t)\,e^{i\omega t}\\
&=2i\sqrt{w\pi}\Theta(\omega)L_{j-1}(2w\omega)e^{-w\omega}\,,
\end{aligned}
\label{eq:app:fourier}
\end{equation}
use $\int_{-\infty}^{+\infty}dx=v_\pm\int_{-\infty}^{+\infty}dt_\pm$ and find
\begin{equation}
\Delta n_{2\pm}(k)=\pm\frac{v_\pm}{2\pi}\sum_{j=1}^{n}|\tilde{\chi}_j(\pm v_\pm k)|^2\,,
\end{equation}
which is precisely the second line of \eqref{eq:nk-stationary}, after substituting the expression \eqref{eq:app:fourier} for $\tilde{\chi}_j(\omega)$.

Let us now focus on the crossed contribution $\Delta n_{2X}(k,t)$, given in \eqref{eq:nkx-gen}. With the help of the representation \eqref{eq:app:chi1}, it can be written as
\begin{equation}
\Delta n_{2X}(k,t)=\int_{-\infty}^{+\infty}\frac{dx}{v_+v_-}\int_{-\infty}^{+\infty}d\xi\,(-i\xi)e^{-ik\xi}\,\mathcal{C}(t,x,\xi)\,,
\end{equation}
with
\begin{equation}
\mathcal{C}(t,x,\xi)=\sum_{r,p=1}^{n}\prod_{\eta=\pm}\chi_r^\eta\left(t_+-\frac{\eta\xi}{2v_+}\right)\chi_p^\eta\left(t_-+\frac{\eta\xi}{2v_-}\right)\,.
\end{equation}
By writing $(-i\xi)e^{-ik\xi}=\partial_ke^{-ik\xi}$ and using the Fourier transform \eqref{eq:app:fourier} we can easily perform the integrations over $x$ and $\xi$ and we readily arrive at the expression \eqref{eq:nk-crossed} given in the main text.

\section{Scaling of the momentum distribution}\label{app:scaling}
In this Appendix we prove the scaling behavior for $k\to 0$ in Eq.\ \eqref{eq:scaling}, obtained in the stationary regime when each fractional excitation can be treated independently from the others. Since the proof is identical for all distributions $\Delta n_{\alpha\pm}$, we focus on $\Delta n_{2+}$:
\begin{equation}
\Delta n_{2+}(k)=\int_{-\infty}^{+\infty}\frac{d\xi}{2\pi}\frac{e^{-ik\xi}}{2\pi i\xi}\underbrace{\int_{-\infty}^{+\infty}dx[e^{i\frac{{q_2}}{q}\,\Delta\varphi_+(x,\xi,t)}-1]}_{\mathcal{I}(\xi)}
\label{eq:app:scaling-start}
\end{equation}
In the limit $k\to 0$ the dominant contribution to the integral $\mathcal{I}(\xi)$ is given by large values of $\xi$. After calculating the phase $\Delta\varphi_+$ for the Lorentzian pulse \eqref{eq:lor-pulse}, the result is
\begin{equation}
\mathcal{I}(\xi)\underset{|\xi|\to\infty}{\approx}|\xi|[e^{i2\pi{q_2}\,\mathrm{sgn}(\xi)}-1]\,,
\label{eq:app:large-xi}
\end{equation}
where we have written the leading term. Finally, we substitute this expression into \eqref{eq:app:scaling-start} and we obtain Eq.\ \eqref{eq:scaling}, with $q_{2+}={q_2}$.

\section{Calculation of the number of excited holes}\label{app:nh}
The purpose of this Appendix is to derive the expression \eqref{eq:nh-T0} for the number of injected holes at zero temperature due to the voltage pulse $V(t)$. Let us start by a simple observation. We defined the momentum distribution for edge channel 2 as
\begin{equation}
\Delta n_2(k,t)=\Braket{c_2^\dagger(k,t)c_2(k,t)-c_2^{0\dagger}(k,t)c_2^0(k,t)}\,,
\end{equation}
with $c_2^0$ the annihilation operator for the equilibrium field $\psi_2$. Now, because of fermionic anticommutation relations we clearly have $\Delta n_2(k,t)=-\Delta\tilde n_2(k,t)$, where
\begin{equation}
\Delta\tilde n_2(k,t)=\Braket{c_2(k,t)c_2^\dagger(k,t)-c_2^0(k,t)c_2^{0\dagger}(k,t)}\,.
\end{equation}
Therefore the number of holes at zero temperature can also be expressed as
\begin{equation}
N_\text{h}(t)=-\int_{-\infty}^{0}dk\,\Delta n_2(k,t)=\int_{-\infty}^{0}dk\,\Delta\tilde n_2(k,t)\,,
\end{equation}
with
\begin{widetext}
	\begin{align}
	\Delta\tilde n_2(k,t)&=\frac{1}{2\pi}\int_{-\infty}^{+\infty}\dif{x}\int_{-\infty}^{+\infty}\dif{\xi}\frac{e^{-ik\xi}}{2\pi}\frac{1}{a-i\xi}\left\{\exp\left[-ie\frac{q_2}{q}\left(\int_{t-\frac{x}{v_+}-\frac{\xi/2}{v_+}}^{t-\frac{x}{v_+}+\frac{\xi/2}{v_+}}\dif{t'}V(t')-\int_{t-\frac{x}{v_-}-\frac{\xi/2}{v_-}}^{t-\frac{x}{v_-}+\frac{\xi/2}{v_-}}\dif{t'}V(t')\right)\right]-1\right\}\,.
	\end{align}
By using the integral representation $\Theta(-k)=\frac{1}{2\pi}\lim_{a\to 0}\int_{-\infty}^{+\infty}dy\,\frac{e^{-iky}}{a+iy}$.
and $\int_{-\infty}^{+\infty}\frac{dy}{(a+iy)^2}=0$, we arrive at
	\begin{equation}
	N_\text{h}(t)=\frac{1}{(2\pi)^2}\int_{-\infty}^{+\infty}dx\int_{-\infty}^{+\infty}dy\frac{1}{(a+iy)^2}\exp\left[-ie\frac{q_2}{q}\left(\int_{t-\frac{x}{v_+}+\frac{y}{2v_+}}^{t-\frac{x}{v_+}-\frac{y}{2v_+}}dt'V(t')-\int_{t-\frac{x}{v_-}+\frac{y}{2v_-}}^{t-\frac{x}{v_-}-\frac{y}{2v_-}}dt'V(t')\right)\right]\,.
	\end{equation}
Furthermore, one can show that the contribution proportional to the sine function in the last expression (we will denote it by $N_\text{h}^\text{s}$) actually vanishes. To that end, we note that
\begin{equation}
\lim_{a\to 0}\frac{1}{(a+iy)^2}=\lim_{a\to 0}\frac{a^2-y^2}{(a^2+y^2)^2}-i\pi\partial_y\delta(y)\equiv A(y)-i\pi\partial_y\delta(y)\,,
\end{equation}
where evidently $A(y)$ is an even function. Therefore $N_\text{h}^{\text{s}}$ becomes
	\begin{align}
	N_\text{h}^\text{s}&=\frac{-i}{(2\pi)^2}\int_{-\infty}^{+\infty}dx\int_{-\infty}^{+\infty}dy[A(y)-i\pi\partial_y\delta(y)]\sin\left[e\frac{q_2}{q}\left(\int_{t-\frac{x}{v_+}+\frac{y}{2v_+}}^{t-\frac{x}{v_+}-\frac{y}{2v_+}}dt'V(t')-\int_{t-\frac{x}{v_-}+\frac{y}{2v_-}}^{t-\frac{x}{v_-}-\frac{y}{2v_-}}dt'V(t')\right)\right]\notag\\
	&=\frac{-i}{(2\pi)^2}\int_{-\infty}^{+\infty}dx\int_{0}^{+\infty}dy[A(y)-A(-y)]\sin\left[e\frac{q_2}{q}\left(\int_{t-\frac{x}{v_+}+\frac{y}{2v_+}}^{t-\frac{x}{v_+}-\frac{y}{2v_+}}dt'V(t')-\int_{t-\frac{x}{v_-}+\frac{y}{2v_-}}^{t-\frac{x}{v_-}-\frac{y}{2v_-}}dt'V(t')\right)\right]\notag\\
	&\quad+\frac{eq_2}{4\pi q}\int_{-\infty}^{+\infty}dx\int_{-\infty}^{+\infty}dy\,\delta(y)\cos\left[e\frac{q_2}{q}\left(\int_{t-\frac{x}{v_+}+\frac{y}{v_+}}^{t-\frac{x}{v_+}}dt'V(t')-\int_{t-\frac{x}{v_-}+\frac{y}{v_-}}^{t-\frac{x}{v_-}}dt'V(t')\right)\right]\sum_{\eta=\pm}\frac{\eta}{v_\eta}V\left(t-\frac{x}{v_\eta}\right)\notag\\
	&=\frac{eq_2}{4\pi q}\int_{-\infty}^{+\infty}dx\left[\frac{1}{v_+}V\left(t-\frac{x}{v_+}\right)-\frac{1}{v_-}V\left(t-\frac{x}{v_-}\right)\right]=0\,,
	\end{align}
\end{widetext}
having used the odd parity of the sine and the even parity of $A(y)$. Equation \eqref{eq:nh-T0} in the main text is thus proved.

\section{Calculation of the excess noise}\label{app:noise}
This Appendix briefly illustrates the calculation of the excess noise \eqref{eq:noise-sp}. The starting point is the definition \eqref{eq:definition-noise}. Up to second order in the tunneling, the time evolution of the current operator $J$ reads
\begin{equation}
\begin{aligned}
J(t)&=J^0(t)-i\int_{t_0}^{t}\dif{\tau}\left[J^0(t),H_\text{T}^0(\tau)\right]+\\
&-\int_{t_0}^{t}\dif{t'}\int_{t_0}^{t'}\dif{t''}\left[H_\text{T}^0(t''),\left[H_\text{T}^0(t'),J^0(t)\right]\right]\,,
\end{aligned}
\end{equation}
the superscript ``0'' denoting the time evolution with respect to $H_\text{edge}+H_\text{V}$. The first term in the previous equation is simply
\begin{equation}
\begin{aligned}
J^0(t)=-\frac{e}{\sqrt{2\pi}}&[v_+\sin\theta\,\partial_x\phi_{L+}(-d+v_+t,0)+\\
&+v_-\cos\theta\,\partial_x\phi_{L-}(-d+v_-t,0)]\,.
\end{aligned}
\end{equation}
First and second order contributions are determined by using the time evolution given by Eqs.\ \eqref{eq:solution-eom-1}, \eqref{eq:solution-eom-2} and \eqref{eq:evolution-Psi}. We obtain:
\begin{align}
S&=4e^2A\int_{-\infty}^{+\infty}dt\int_{-\infty}^{+\infty}d\tau\left(\frac{1}{a+iv_-\tau}\frac{\pi T\tau}{\sinh\pi T\tau}\right)^2\notag\\
&\times\cos\left[e\frac{q_2}{q}\left(\int_{t+\tau}^{t}dt'V(t')-\int_{t+\tau+\tau_d}^{t+\tau_d}dt'V(t')\right)\right]\label{eq:app:noise}
\end{align}
\begin{align}
\overline{\Braket{J(t)}}&=-2ieA\int_{-\infty}^{+\infty}dt\int_{-\infty}^{+\infty}d\tau\left(\frac{1}{a+iv_-\tau}\frac{\pi T\tau}{\sinh\pi T\tau}\right)^2\notag\\
&\times\sin\left[e\frac{q_2}{q}\left(\int_{t-\tau}^{t}dt'V(t')-\int_{t-\tau+\tau_d}^{t+\tau_d}dt'V(t')\right)\right]\,,
\end{align}
where $\gamma=v_-/v_+$ and $A=|\Lambda|^2\gamma^{2\sin^2\!\theta}(2\pi)^{-2}$.
Now, by using the same procedure as in App.\ \ref{app:nh}, we can write the current as
\begin{equation}
\overline{\Braket{J(t)}}=\frac{2\pi e^2q_2A}{qv_-^2}\int_{-\infty}^{+\infty}dt[V(t)-V(t+\tau_d)]=0\,.
\end{equation}
Finally, we note that the constant $\gamma^{2\sin^2\!\theta}$ appearing in $A$ can be reabsorbed in the tunneling amplitude $\Lambda$, so that Eq.\ \eqref{eq:app:noise} becomes identical to Eq.\ \eqref{eq:noise-sp} given in the main text.

\bibliography{refsEQO}

\begin{thebibliography}{91}%
\makeatletter
\providecommand \@ifxundefined [1]{%
 \@ifx{#1\undefined}
}%
\providecommand \@ifnum [1]{%
 \ifnum #1\expandafter \@firstoftwo
 \else \expandafter \@secondoftwo
 \fi
}%
\providecommand \@ifx [1]{%
 \ifx #1\expandafter \@firstoftwo
 \else \expandafter \@secondoftwo
 \fi
}%
\providecommand \natexlab [1]{#1}%
\providecommand \enquote  [1]{``#1''}%
\providecommand \bibnamefont  [1]{#1}%
\providecommand \bibfnamefont [1]{#1}%
\providecommand \citenamefont [1]{#1}%
\providecommand \href@noop [0]{\@secondoftwo}%
\providecommand \href [0]{\begingroup \@sanitize@url \@href}%
\providecommand \@href[1]{\@@startlink{#1}\@@href}%
\providecommand \@@href[1]{\endgroup#1\@@endlink}%
\providecommand \@sanitize@url [0]{\catcode `\\12\catcode `\$12\catcode
  `\&12\catcode `\#12\catcode `\^12\catcode `\_12\catcode `\%12\relax}%
\providecommand \@@startlink[1]{}%
\providecommand \@@endlink[0]{}%
\providecommand \url  [0]{\begingroup\@sanitize@url \@url }%
\providecommand \@url [1]{\endgroup\@href {#1}{\urlprefix }}%
\providecommand \urlprefix  [0]{URL }%
\providecommand \Eprint [0]{\href }%
\providecommand \doibase [0]{http://dx.doi.org/}%
\providecommand \selectlanguage [0]{\@gobble}%
\providecommand \bibinfo  [0]{\@secondoftwo}%
\providecommand \bibfield  [0]{\@secondoftwo}%
\providecommand \translation [1]{[#1]}%
\providecommand \BibitemOpen [0]{}%
\providecommand \bibitemStop [0]{}%
\providecommand \bibitemNoStop [0]{.\EOS\space}%
\providecommand \EOS [0]{\spacefactor3000\relax}%
\providecommand \BibitemShut  [1]{\csname bibitem#1\endcsname}%
\let\auto@bib@innerbib\@empty
\bibitem [{\citenamefont {Klitzing}\ \emph {et~al.}(1980)\citenamefont
  {Klitzing}, \citenamefont {Dorda},\ and\ \citenamefont
  {Pepper}}]{klitzing80}%
  \BibitemOpen
  \bibfield  {author} {\bibinfo {author} {\bibfnamefont {K.~v.}\ \bibnamefont
  {Klitzing}}, \bibinfo {author} {\bibfnamefont {G.}~\bibnamefont {Dorda}}, \
  and\ \bibinfo {author} {\bibfnamefont {M.}~\bibnamefont {Pepper}},\ }\href
  {\doibase 10.1103/PhysRevLett.45.494} {\bibfield  {journal} {\bibinfo
  {journal} {Phys. Rev. Lett.}\ }\textbf {\bibinfo {volume} {45}},\ \bibinfo
  {pages} {494} (\bibinfo {year} {1980})}\BibitemShut {NoStop}%
\bibitem [{\citenamefont {Prange}\ and\ \citenamefont
  {Girvin}(1990)}]{girvin1990}%
  \BibitemOpen
  \bibfield  {author} {\bibinfo {author} {\bibfnamefont {R.~E.}\ \bibnamefont
  {Prange}}\ and\ \bibinfo {author} {\bibfnamefont {S.~M.}\ \bibnamefont
  {Girvin}},\ }\href@noop {} {\emph {\bibinfo {title} {The Quantum Hall
  Effect}}}\ (\bibinfo  {publisher} {Springer-Verlag},\ \bibinfo {year}
  {1990})\BibitemShut {NoStop}%
\bibitem [{\citenamefont {{Girvin}}(1999)}]{girvin99}%
  \BibitemOpen
  \bibfield  {author} {\bibinfo {author} {\bibfnamefont {S.~M.}\ \bibnamefont
  {{Girvin}}},\ }in\ \href@noop {} {\emph {\bibinfo {booktitle} {Topological
  Aspects of Low Dimensional Systems}}},\ \bibinfo {editor} {edited by\
  \bibinfo {editor} {\bibfnamefont {A.}~\bibnamefont {{Comtet}}}, \bibinfo
  {editor} {\bibfnamefont {T.}~\bibnamefont {{Jolicoeur}}}, \bibinfo {editor}
  {\bibfnamefont {S.}~\bibnamefont {{Ouvry}}}, \ and\ \bibinfo {editor}
  {\bibfnamefont {F.}~\bibnamefont {{David}}}}\ (\bibinfo {year} {1999})\
  p.~\bibinfo {pages} {53},\ \Eprint {http://arxiv.org/abs/cond-mat/9907002}
  {cond-mat/9907002} \BibitemShut {NoStop}%
\bibitem [{\citenamefont {Stern}(2008)}]{stern-review}%
  \BibitemOpen
  \bibfield  {author} {\bibinfo {author} {\bibfnamefont {A.}~\bibnamefont
  {Stern}},\ }\href {\doibase https://doi.org/10.1016/j.aop.2007.10.008}
  {\bibfield  {journal} {\bibinfo  {journal} {Annals of Physics}\ }\textbf
  {\bibinfo {volume} {323}},\ \bibinfo {pages} {204} (\bibinfo {year}
  {2008})}\BibitemShut {NoStop}%
\bibitem [{\citenamefont {{Goerbig}}()}]{goerbig09}%
  \BibitemOpen
  \bibfield  {author} {\bibinfo {author} {\bibfnamefont {M.~O.}\ \bibnamefont
  {{Goerbig}}},\ }\href@noop {} {\ }\Eprint {http://arxiv.org/abs/0909.1998}
  {arXiv:0909.1998} \BibitemShut {NoStop}%
\bibitem [{\citenamefont {Grenier}\ \emph {et~al.}(2011)\citenamefont
  {Grenier}, \citenamefont {Herv\'e}, \citenamefont {F\`eve},\ and\
  \citenamefont {Degiovanni}}]{grenier11}%
  \BibitemOpen
  \bibfield  {author} {\bibinfo {author} {\bibfnamefont {C.}~\bibnamefont
  {Grenier}}, \bibinfo {author} {\bibfnamefont {R.}~\bibnamefont {Herv\'e}},
  \bibinfo {author} {\bibfnamefont {G.}~\bibnamefont {F\`eve}}, \ and\ \bibinfo
  {author} {\bibfnamefont {P.}~\bibnamefont {Degiovanni}},\ }\href {\doibase
  10.1142/S0217984911026772} {\bibfield  {journal} {\bibinfo  {journal} {Modern
  Physics Letters B}\ }\textbf {\bibinfo {volume} {25}},\ \bibinfo {pages}
  {1053} (\bibinfo {year} {2011})}\BibitemShut {NoStop}%
\bibitem [{\citenamefont {Bocquillon}\ \emph {et~al.}(2014)\citenamefont
  {Bocquillon}, \citenamefont {Freulon}, \citenamefont {Parmentier},
  \citenamefont {Berroir}, \citenamefont {Pla\c{c}ais}, \citenamefont {Wahl},
  \citenamefont {Rech}, \citenamefont {Jonckheere}, \citenamefont {Martin},
  \citenamefont {Grenier}, \citenamefont {Ferraro}, \citenamefont
  {Degiovanni},\ and\ \citenamefont {F\`eve}}]{bocquillon14}%
  \BibitemOpen
  \bibfield  {author} {\bibinfo {author} {\bibfnamefont {E.}~\bibnamefont
  {Bocquillon}}, \bibinfo {author} {\bibfnamefont {V.}~\bibnamefont {Freulon}},
  \bibinfo {author} {\bibfnamefont {F.~D.}\ \bibnamefont {Parmentier}},
  \bibinfo {author} {\bibfnamefont {J.}~\bibnamefont {Berroir}}, \bibinfo
  {author} {\bibfnamefont {B.}~\bibnamefont {Pla\c{c}ais}}, \bibinfo {author}
  {\bibfnamefont {C.}~\bibnamefont {Wahl}}, \bibinfo {author} {\bibfnamefont
  {J.}~\bibnamefont {Rech}}, \bibinfo {author} {\bibfnamefont {T.}~\bibnamefont
  {Jonckheere}}, \bibinfo {author} {\bibfnamefont {T.}~\bibnamefont {Martin}},
  \bibinfo {author} {\bibfnamefont {C.}~\bibnamefont {Grenier}}, \bibinfo
  {author} {\bibfnamefont {D.}~\bibnamefont {Ferraro}}, \bibinfo {author}
  {\bibfnamefont {P.}~\bibnamefont {Degiovanni}}, \ and\ \bibinfo {author}
  {\bibfnamefont {G.}~\bibnamefont {F\`eve}},\ }\href {\doibase
  10.1002/andp.201300181} {\bibfield  {journal} {\bibinfo  {journal} {Annalen
  der Physik}\ }\textbf {\bibinfo {volume} {526}},\ \bibinfo {pages} {1}
  (\bibinfo {year} {2014})}\BibitemShut {NoStop}%
\bibitem [{\citenamefont {Haldane}(1988)}]{haldane1988prl}%
  \BibitemOpen
  \bibfield  {author} {\bibinfo {author} {\bibfnamefont {F.~D.~M.}\
  \bibnamefont {Haldane}},\ }\href {\doibase 10.1103/PhysRevLett.61.2015}
  {\bibfield  {journal} {\bibinfo  {journal} {Phys. Rev. Lett.}\ }\textbf
  {\bibinfo {volume} {61}},\ \bibinfo {pages} {2015} (\bibinfo {year}
  {1988})}\BibitemShut {NoStop}%
\bibitem [{\citenamefont {K{\"o}nig}\ \emph {et~al.}(2007)\citenamefont
  {K{\"o}nig}, \citenamefont {Wiedmann}, \citenamefont {Br{\"u}ne},
  \citenamefont {Roth}, \citenamefont {Buhmann}, \citenamefont {Molenkamp},
  \citenamefont {Qi},\ and\ \citenamefont {Zhang}}]{konig07qsh}%
  \BibitemOpen
  \bibfield  {author} {\bibinfo {author} {\bibfnamefont {M.}~\bibnamefont
  {K{\"o}nig}}, \bibinfo {author} {\bibfnamefont {S.}~\bibnamefont {Wiedmann}},
  \bibinfo {author} {\bibfnamefont {C.}~\bibnamefont {Br{\"u}ne}}, \bibinfo
  {author} {\bibfnamefont {A.}~\bibnamefont {Roth}}, \bibinfo {author}
  {\bibfnamefont {H.}~\bibnamefont {Buhmann}}, \bibinfo {author} {\bibfnamefont
  {L.~W.}\ \bibnamefont {Molenkamp}}, \bibinfo {author} {\bibfnamefont {X.-L.}\
  \bibnamefont {Qi}}, \ and\ \bibinfo {author} {\bibfnamefont {S.-C.}\
  \bibnamefont {Zhang}},\ }\href {\doibase 10.1126/science.1148047} {\bibfield
  {journal} {\bibinfo  {journal} {Science}\ }\textbf {\bibinfo {volume}
  {318}},\ \bibinfo {pages} {766} (\bibinfo {year} {2007})}\BibitemShut
  {NoStop}%
\bibitem [{\citenamefont {Hasan}\ and\ \citenamefont
  {Kane}(2010)}]{hasan2010colloquium}%
  \BibitemOpen
  \bibfield  {author} {\bibinfo {author} {\bibfnamefont {M.~Z.}\ \bibnamefont
  {Hasan}}\ and\ \bibinfo {author} {\bibfnamefont {C.~L.}\ \bibnamefont
  {Kane}},\ }\href {\doibase 10.1103/RevModPhys.82.3045} {\bibfield  {journal}
  {\bibinfo  {journal} {Rev. Mod. Phys.}\ }\textbf {\bibinfo {volume} {82}},\
  \bibinfo {pages} {3045} (\bibinfo {year} {2010})}\BibitemShut {NoStop}%
\bibitem [{\citenamefont {Hofer}\ and\ \citenamefont
  {B\"uttiker}(2013)}]{hofer2013}%
  \BibitemOpen
  \bibfield  {author} {\bibinfo {author} {\bibfnamefont {P.~P.}\ \bibnamefont
  {Hofer}}\ and\ \bibinfo {author} {\bibfnamefont {M.}~\bibnamefont
  {B\"uttiker}},\ }\href {\doibase 10.1103/PhysRevB.88.241308} {\bibfield
  {journal} {\bibinfo  {journal} {Phys. Rev. B}\ }\textbf {\bibinfo {volume}
  {88}},\ \bibinfo {pages} {241308} (\bibinfo {year} {2013})}\BibitemShut
  {NoStop}%
\bibitem [{\citenamefont {Inhofer}\ and\ \citenamefont
  {Bercioux}(2013)}]{inhofer2013}%
  \BibitemOpen
  \bibfield  {author} {\bibinfo {author} {\bibfnamefont {A.}~\bibnamefont
  {Inhofer}}\ and\ \bibinfo {author} {\bibfnamefont {D.}~\bibnamefont
  {Bercioux}},\ }\href {\doibase 10.1103/PhysRevB.88.235412} {\bibfield
  {journal} {\bibinfo  {journal} {Phys. Rev. B}\ }\textbf {\bibinfo {volume}
  {88}},\ \bibinfo {pages} {235412} (\bibinfo {year} {2013})}\BibitemShut
  {NoStop}%
\bibitem [{\citenamefont {Ferraro}\ \emph
  {et~al.}(2014{\natexlab{a}})\citenamefont {Ferraro}, \citenamefont {Wahl},
  \citenamefont {Rech}, \citenamefont {Jonckheere},\ and\ \citenamefont
  {Martin}}]{ferraro14HOMtopo}%
  \BibitemOpen
  \bibfield  {author} {\bibinfo {author} {\bibfnamefont {D.}~\bibnamefont
  {Ferraro}}, \bibinfo {author} {\bibfnamefont {C.}~\bibnamefont {Wahl}},
  \bibinfo {author} {\bibfnamefont {J.}~\bibnamefont {Rech}}, \bibinfo {author}
  {\bibfnamefont {T.}~\bibnamefont {Jonckheere}}, \ and\ \bibinfo {author}
  {\bibfnamefont {T.}~\bibnamefont {Martin}},\ }\href {\doibase
  10.1103/PhysRevB.89.075407} {\bibfield  {journal} {\bibinfo  {journal} {Phys.
  Rev. B}\ }\textbf {\bibinfo {volume} {89}},\ \bibinfo {pages} {075407}
  (\bibinfo {year} {2014}{\natexlab{a}})}\BibitemShut {NoStop}%
\bibitem [{\citenamefont {Str\"om}\ \emph {et~al.}(2015)\citenamefont
  {Str\"om}, \citenamefont {Johannesson},\ and\ \citenamefont
  {Recher}}]{strom15entanglement}%
  \BibitemOpen
  \bibfield  {author} {\bibinfo {author} {\bibfnamefont {A.}~\bibnamefont
  {Str\"om}}, \bibinfo {author} {\bibfnamefont {H.}~\bibnamefont
  {Johannesson}}, \ and\ \bibinfo {author} {\bibfnamefont {P.}~\bibnamefont
  {Recher}},\ }\href {\doibase 10.1103/PhysRevB.91.245406} {\bibfield
  {journal} {\bibinfo  {journal} {Phys. Rev. B}\ }\textbf {\bibinfo {volume}
  {91}},\ \bibinfo {pages} {245406} (\bibinfo {year} {2015})}\BibitemShut
  {NoStop}%
\bibitem [{\citenamefont {Dolcini}\ \emph {et~al.}(2016)\citenamefont
  {Dolcini}, \citenamefont {Iotti}, \citenamefont {Montorsi},\ and\
  \citenamefont {Rossi}}]{dolcini16chiral}%
  \BibitemOpen
  \bibfield  {author} {\bibinfo {author} {\bibfnamefont {F.}~\bibnamefont
  {Dolcini}}, \bibinfo {author} {\bibfnamefont {R.~C.}\ \bibnamefont {Iotti}},
  \bibinfo {author} {\bibfnamefont {A.}~\bibnamefont {Montorsi}}, \ and\
  \bibinfo {author} {\bibfnamefont {F.}~\bibnamefont {Rossi}},\ }\href
  {\doibase 10.1103/PhysRevB.94.165412} {\bibfield  {journal} {\bibinfo
  {journal} {Phys. Rev. B}\ }\textbf {\bibinfo {volume} {94}},\ \bibinfo
  {pages} {165412} (\bibinfo {year} {2016})}\BibitemShut {NoStop}%
\bibitem [{\citenamefont {Calzona}\ \emph {et~al.}(2016)\citenamefont
  {Calzona}, \citenamefont {Acciai}, \citenamefont {Carrega}, \citenamefont
  {Cavaliere},\ and\ \citenamefont {Sassetti}}]{calzona16energypart}%
  \BibitemOpen
  \bibfield  {author} {\bibinfo {author} {\bibfnamefont {A.}~\bibnamefont
  {Calzona}}, \bibinfo {author} {\bibfnamefont {M.}~\bibnamefont {Acciai}},
  \bibinfo {author} {\bibfnamefont {M.}~\bibnamefont {Carrega}}, \bibinfo
  {author} {\bibfnamefont {F.}~\bibnamefont {Cavaliere}}, \ and\ \bibinfo
  {author} {\bibfnamefont {M.}~\bibnamefont {Sassetti}},\ }\href {\doibase
  10.1103/PhysRevB.94.035404} {\bibfield  {journal} {\bibinfo  {journal} {Phys.
  Rev. B}\ }\textbf {\bibinfo {volume} {94}},\ \bibinfo {pages} {035404}
  (\bibinfo {year} {2016})}\BibitemShut {NoStop}%
\bibitem [{\citenamefont {Acciai}\ \emph {et~al.}(2017)\citenamefont {Acciai},
  \citenamefont {Calzona}, \citenamefont {Dolcetto}, \citenamefont {Schmidt},\
  and\ \citenamefont {Sassetti}}]{acciai17}%
  \BibitemOpen
  \bibfield  {author} {\bibinfo {author} {\bibfnamefont {M.}~\bibnamefont
  {Acciai}}, \bibinfo {author} {\bibfnamefont {A.}~\bibnamefont {Calzona}},
  \bibinfo {author} {\bibfnamefont {G.}~\bibnamefont {Dolcetto}}, \bibinfo
  {author} {\bibfnamefont {T.~L.}\ \bibnamefont {Schmidt}}, \ and\ \bibinfo
  {author} {\bibfnamefont {M.}~\bibnamefont {Sassetti}},\ }\href {\doibase
  10.1103/PhysRevB.96.075144} {\bibfield  {journal} {\bibinfo  {journal} {Phys.
  Rev. B}\ }\textbf {\bibinfo {volume} {96}},\ \bibinfo {pages} {075144}
  (\bibinfo {year} {2017})}\BibitemShut {NoStop}%
\bibitem [{\citenamefont {F{\`e}ve}\ \emph {et~al.}(2007)\citenamefont
  {F{\`e}ve}, \citenamefont {Mah{\'e}}, \citenamefont {Berroir}, \citenamefont
  {Kontos}, \citenamefont {Pla{\c c}ais}, \citenamefont {Glattli},
  \citenamefont {Cavanna}, \citenamefont {Etienne},\ and\ \citenamefont
  {Jin}}]{feve07}%
  \BibitemOpen
  \bibfield  {author} {\bibinfo {author} {\bibfnamefont {G.}~\bibnamefont
  {F{\`e}ve}}, \bibinfo {author} {\bibfnamefont {A.}~\bibnamefont {Mah{\'e}}},
  \bibinfo {author} {\bibfnamefont {J.-M.}\ \bibnamefont {Berroir}}, \bibinfo
  {author} {\bibfnamefont {T.}~\bibnamefont {Kontos}}, \bibinfo {author}
  {\bibfnamefont {B.}~\bibnamefont {Pla{\c c}ais}}, \bibinfo {author}
  {\bibfnamefont {D.~C.}\ \bibnamefont {Glattli}}, \bibinfo {author}
  {\bibfnamefont {A.}~\bibnamefont {Cavanna}}, \bibinfo {author} {\bibfnamefont
  {B.}~\bibnamefont {Etienne}}, \ and\ \bibinfo {author} {\bibfnamefont
  {Y.}~\bibnamefont {Jin}},\ }\href {\doibase 10.1126/science.1141243}
  {\bibfield  {journal} {\bibinfo  {journal} {Science}\ }\textbf {\bibinfo
  {volume} {316}},\ \bibinfo {pages} {1169} (\bibinfo {year}
  {2007})}\BibitemShut {NoStop}%
\bibitem [{\citenamefont {Moskalets}\ \emph {et~al.}(2008)\citenamefont
  {Moskalets}, \citenamefont {Samuelsson},\ and\ \citenamefont
  {B\"uttiker}}]{moskalets08meso}%
  \BibitemOpen
  \bibfield  {author} {\bibinfo {author} {\bibfnamefont {M.}~\bibnamefont
  {Moskalets}}, \bibinfo {author} {\bibfnamefont {P.}~\bibnamefont
  {Samuelsson}}, \ and\ \bibinfo {author} {\bibfnamefont {M.}~\bibnamefont
  {B\"uttiker}},\ }\href {\doibase 10.1103/PhysRevLett.100.086601} {\bibfield
  {journal} {\bibinfo  {journal} {Phys. Rev. Lett.}\ }\textbf {\bibinfo
  {volume} {100}},\ \bibinfo {pages} {086601} (\bibinfo {year}
  {2008})}\BibitemShut {NoStop}%
\bibitem [{\citenamefont {Mah\'e}\ \emph {et~al.}(2010)\citenamefont {Mah\'e},
  \citenamefont {Parmentier}, \citenamefont {Bocquillon}, \citenamefont
  {Berroir}, \citenamefont {Glattli}, \citenamefont {Kontos}, \citenamefont
  {Pla\ifmmode~\mbox{\c{c}}\else \c{c}\fi{}ais}, \citenamefont {F\`eve},
  \citenamefont {Cavanna},\ and\ \citenamefont {Jin}}]{mahe2010}%
  \BibitemOpen
  \bibfield  {author} {\bibinfo {author} {\bibfnamefont {A.}~\bibnamefont
  {Mah\'e}}, \bibinfo {author} {\bibfnamefont {F.~D.}\ \bibnamefont
  {Parmentier}}, \bibinfo {author} {\bibfnamefont {E.}~\bibnamefont
  {Bocquillon}}, \bibinfo {author} {\bibfnamefont {J.-M.}\ \bibnamefont
  {Berroir}}, \bibinfo {author} {\bibfnamefont {D.~C.}\ \bibnamefont
  {Glattli}}, \bibinfo {author} {\bibfnamefont {T.}~\bibnamefont {Kontos}},
  \bibinfo {author} {\bibfnamefont {B.}~\bibnamefont
  {Pla\ifmmode~\mbox{\c{c}}\else \c{c}\fi{}ais}}, \bibinfo {author}
  {\bibfnamefont {G.}~\bibnamefont {F\`eve}}, \bibinfo {author} {\bibfnamefont
  {A.}~\bibnamefont {Cavanna}}, \ and\ \bibinfo {author} {\bibfnamefont
  {Y.}~\bibnamefont {Jin}},\ }\href {\doibase 10.1103/PhysRevB.82.201309}
  {\bibfield  {journal} {\bibinfo  {journal} {Phys. Rev. B}\ }\textbf {\bibinfo
  {volume} {82}},\ \bibinfo {pages} {201309} (\bibinfo {year}
  {2010})}\BibitemShut {NoStop}%
\bibitem [{\citenamefont {Dubois}\ \emph
  {et~al.}(2013{\natexlab{a}})\citenamefont {Dubois}, \citenamefont {Jullien},
  \citenamefont {Portier}, \citenamefont {Roche}, \citenamefont {Cavanna},
  \citenamefont {Jin}, \citenamefont {Wegscheider}, \citenamefont {Roulleau},\
  and\ \citenamefont {Glattli}}]{dubois2013levitonsNature}%
  \BibitemOpen
  \bibfield  {author} {\bibinfo {author} {\bibfnamefont {J.}~\bibnamefont
  {Dubois}}, \bibinfo {author} {\bibfnamefont {T.}~\bibnamefont {Jullien}},
  \bibinfo {author} {\bibfnamefont {F.}~\bibnamefont {Portier}}, \bibinfo
  {author} {\bibfnamefont {P.}~\bibnamefont {Roche}}, \bibinfo {author}
  {\bibfnamefont {A.}~\bibnamefont {Cavanna}}, \bibinfo {author} {\bibfnamefont
  {Y.}~\bibnamefont {Jin}}, \bibinfo {author} {\bibfnamefont {W.}~\bibnamefont
  {Wegscheider}}, \bibinfo {author} {\bibfnamefont {P.}~\bibnamefont
  {Roulleau}}, \ and\ \bibinfo {author} {\bibfnamefont {D.~C.}\ \bibnamefont
  {Glattli}},\ }\href {\doibase 10.1038/nature12713} {\bibfield  {journal}
  {\bibinfo  {journal} {Nature}\ }\textbf {\bibinfo {volume} {502}},\ \bibinfo
  {pages} {659} (\bibinfo {year} {2013}{\natexlab{a}})}\BibitemShut {NoStop}%
\bibitem [{\citenamefont {Jullien}\ \emph {et~al.}(2014)\citenamefont
  {Jullien}, \citenamefont {Roulleau}, \citenamefont {Roche}, \citenamefont
  {Cavanna}, \citenamefont {Jin},\ and\ \citenamefont
  {Glattli}}]{jullien14tomography}%
  \BibitemOpen
  \bibfield  {author} {\bibinfo {author} {\bibfnamefont {T.}~\bibnamefont
  {Jullien}}, \bibinfo {author} {\bibfnamefont {P.}~\bibnamefont {Roulleau}},
  \bibinfo {author} {\bibfnamefont {B.}~\bibnamefont {Roche}}, \bibinfo
  {author} {\bibfnamefont {A.}~\bibnamefont {Cavanna}}, \bibinfo {author}
  {\bibfnamefont {Y.}~\bibnamefont {Jin}}, \ and\ \bibinfo {author}
  {\bibfnamefont {D.~C.}\ \bibnamefont {Glattli}},\ }\href {\doibase
  10.1038/nature13821} {\bibfield  {journal} {\bibinfo  {journal} {Nature}\
  }\textbf {\bibinfo {volume} {514}},\ \bibinfo {pages} {603} (\bibinfo {year}
  {2014})}\BibitemShut {NoStop}%
\bibitem [{\citenamefont {Levitov}\ \emph {et~al.}(1996)\citenamefont
  {Levitov}, \citenamefont {Lee},\ and\ \citenamefont {Lesovik}}]{levitov96}%
  \BibitemOpen
  \bibfield  {author} {\bibinfo {author} {\bibfnamefont {L.~S.}\ \bibnamefont
  {Levitov}}, \bibinfo {author} {\bibfnamefont {H.}~\bibnamefont {Lee}}, \ and\
  \bibinfo {author} {\bibfnamefont {G.~B.}\ \bibnamefont {Lesovik}},\ }\href
  {\doibase 10.1063/1.531672} {\bibfield  {journal} {\bibinfo  {journal}
  {Journal of Mathematical Physics}\ }\textbf {\bibinfo {volume} {37}},\
  \bibinfo {pages} {4845} (\bibinfo {year} {1996})}\BibitemShut {NoStop}%
\bibitem [{\citenamefont {Ivanov}\ \emph {et~al.}(1997)\citenamefont {Ivanov},
  \citenamefont {Lee},\ and\ \citenamefont {Levitov}}]{levitov97}%
  \BibitemOpen
  \bibfield  {author} {\bibinfo {author} {\bibfnamefont {D.~A.}\ \bibnamefont
  {Ivanov}}, \bibinfo {author} {\bibfnamefont {H.~W.}\ \bibnamefont {Lee}}, \
  and\ \bibinfo {author} {\bibfnamefont {L.~S.}\ \bibnamefont {Levitov}},\
  }\href {\doibase 10.1103/PhysRevB.56.6839} {\bibfield  {journal} {\bibinfo
  {journal} {Phys. Rev. B}\ }\textbf {\bibinfo {volume} {56}},\ \bibinfo
  {pages} {6839} (\bibinfo {year} {1997})}\BibitemShut {NoStop}%
\bibitem [{\citenamefont {Bocquillon}\ \emph {et~al.}(2012)\citenamefont
  {Bocquillon}, \citenamefont {Parmentier}, \citenamefont {Grenier},
  \citenamefont {Berroir}, \citenamefont {Degiovanni}, \citenamefont {Glattli},
  \citenamefont {Pla\ifmmode~\mbox{\c{c}}\else \c{c}\fi{}ais}, \citenamefont
  {Cavanna}, \citenamefont {Jin},\ and\ \citenamefont
  {F\`eve}}]{bocquillon12eqo}%
  \BibitemOpen
  \bibfield  {author} {\bibinfo {author} {\bibfnamefont {E.}~\bibnamefont
  {Bocquillon}}, \bibinfo {author} {\bibfnamefont {F.~D.}\ \bibnamefont
  {Parmentier}}, \bibinfo {author} {\bibfnamefont {C.}~\bibnamefont {Grenier}},
  \bibinfo {author} {\bibfnamefont {J.-M.}\ \bibnamefont {Berroir}}, \bibinfo
  {author} {\bibfnamefont {P.}~\bibnamefont {Degiovanni}}, \bibinfo {author}
  {\bibfnamefont {D.~C.}\ \bibnamefont {Glattli}}, \bibinfo {author}
  {\bibfnamefont {B.}~\bibnamefont {Pla\ifmmode~\mbox{\c{c}}\else
  \c{c}\fi{}ais}}, \bibinfo {author} {\bibfnamefont {A.}~\bibnamefont
  {Cavanna}}, \bibinfo {author} {\bibfnamefont {Y.}~\bibnamefont {Jin}}, \ and\
  \bibinfo {author} {\bibfnamefont {G.}~\bibnamefont {F\`eve}},\ }\href
  {\doibase 10.1103/PhysRevLett.108.196803} {\bibfield  {journal} {\bibinfo
  {journal} {Phys. Rev. Lett.}\ }\textbf {\bibinfo {volume} {108}},\ \bibinfo
  {pages} {196803} (\bibinfo {year} {2012})}\BibitemShut {NoStop}%
\bibitem [{\citenamefont {Bocquillon}\ \emph
  {et~al.}(2013{\natexlab{a}})\citenamefont {Bocquillon}, \citenamefont
  {Freulon}, \citenamefont {Berroir}, \citenamefont {Degiovanni}, \citenamefont
  {Pla{\c c}ais}, \citenamefont {Cavanna}, \citenamefont {Jin},\ and\
  \citenamefont {F{\`e}ve}}]{bocquillon13homscience}%
  \BibitemOpen
  \bibfield  {author} {\bibinfo {author} {\bibfnamefont {E.}~\bibnamefont
  {Bocquillon}}, \bibinfo {author} {\bibfnamefont {V.}~\bibnamefont {Freulon}},
  \bibinfo {author} {\bibfnamefont {J.-M.}\ \bibnamefont {Berroir}}, \bibinfo
  {author} {\bibfnamefont {P.}~\bibnamefont {Degiovanni}}, \bibinfo {author}
  {\bibfnamefont {B.}~\bibnamefont {Pla{\c c}ais}}, \bibinfo {author}
  {\bibfnamefont {A.}~\bibnamefont {Cavanna}}, \bibinfo {author} {\bibfnamefont
  {Y.}~\bibnamefont {Jin}}, \ and\ \bibinfo {author} {\bibfnamefont
  {G.}~\bibnamefont {F{\`e}ve}},\ }\href {\doibase 10.1126/science.1232572}
  {\bibfield  {journal} {\bibinfo  {journal} {Science}\ }\textbf {\bibinfo
  {volume} {339}},\ \bibinfo {pages} {1054} (\bibinfo {year}
  {2013}{\natexlab{a}})}\BibitemShut {NoStop}%
\bibitem [{\citenamefont {Freulon}\ \emph {et~al.}(2015)\citenamefont
  {Freulon}, \citenamefont {Marguerite}, \citenamefont {Berroir}, \citenamefont
  {Pla{\c c}ais}, \citenamefont {Cavanna}, \citenamefont {Jin},\ and\
  \citenamefont {F\`eve}}]{freulon15hom}%
  \BibitemOpen
  \bibfield  {author} {\bibinfo {author} {\bibfnamefont {V.}~\bibnamefont
  {Freulon}}, \bibinfo {author} {\bibfnamefont {A.}~\bibnamefont {Marguerite}},
  \bibinfo {author} {\bibfnamefont {J.-M.}\ \bibnamefont {Berroir}}, \bibinfo
  {author} {\bibfnamefont {B.}~\bibnamefont {Pla{\c c}ais}}, \bibinfo {author}
  {\bibfnamefont {A.}~\bibnamefont {Cavanna}}, \bibinfo {author} {\bibfnamefont
  {Y.}~\bibnamefont {Jin}}, \ and\ \bibinfo {author} {\bibfnamefont
  {G.}~\bibnamefont {F\`eve}},\ }\href {\doibase 10.1038/ncomms7854} {\bibfield
   {journal} {\bibinfo  {journal} {Nature Communications}\ }\textbf {\bibinfo
  {volume} {6}},\ \bibinfo {pages} {6854} (\bibinfo {year} {2015})}\BibitemShut
  {NoStop}%
\bibitem [{\citenamefont {Tewari}\ \emph {et~al.}(2016)\citenamefont {Tewari},
  \citenamefont {Roulleau}, \citenamefont {Grenier}, \citenamefont {Portier},
  \citenamefont {Cavanna}, \citenamefont {Gennser}, \citenamefont {Mailly},\
  and\ \citenamefont {Roche}}]{tewari2016}%
  \BibitemOpen
  \bibfield  {author} {\bibinfo {author} {\bibfnamefont {S.}~\bibnamefont
  {Tewari}}, \bibinfo {author} {\bibfnamefont {P.}~\bibnamefont {Roulleau}},
  \bibinfo {author} {\bibfnamefont {C.}~\bibnamefont {Grenier}}, \bibinfo
  {author} {\bibfnamefont {F.}~\bibnamefont {Portier}}, \bibinfo {author}
  {\bibfnamefont {A.}~\bibnamefont {Cavanna}}, \bibinfo {author} {\bibfnamefont
  {U.}~\bibnamefont {Gennser}}, \bibinfo {author} {\bibfnamefont
  {D.}~\bibnamefont {Mailly}}, \ and\ \bibinfo {author} {\bibfnamefont
  {P.}~\bibnamefont {Roche}},\ }\href {\doibase 10.1103/PhysRevB.93.035420}
  {\bibfield  {journal} {\bibinfo  {journal} {Phys. Rev. B}\ }\textbf {\bibinfo
  {volume} {93}},\ \bibinfo {pages} {035420} (\bibinfo {year}
  {2016})}\BibitemShut {NoStop}%
\bibitem [{\citenamefont {Hanbury~Brown}\ and\ \citenamefont
  {Twiss}(1956)}]{hbt56}%
  \BibitemOpen
  \bibfield  {author} {\bibinfo {author} {\bibfnamefont {R.}~\bibnamefont
  {Hanbury~Brown}}\ and\ \bibinfo {author} {\bibfnamefont {R.~Q.}\ \bibnamefont
  {Twiss}},\ }\href {\doibase 10.1038/1781046a0} {\bibfield  {journal}
  {\bibinfo  {journal} {Nature}\ }\textbf {\bibinfo {volume} {178}},\ \bibinfo
  {pages} {1046} (\bibinfo {year} {1956})}\BibitemShut {NoStop}%
\bibitem [{\citenamefont {Hong}\ \emph {et~al.}(1987)\citenamefont {Hong},
  \citenamefont {Ou},\ and\ \citenamefont {Mandel}}]{hom87}%
  \BibitemOpen
  \bibfield  {author} {\bibinfo {author} {\bibfnamefont {C.~K.}\ \bibnamefont
  {Hong}}, \bibinfo {author} {\bibfnamefont {Z.~Y.}\ \bibnamefont {Ou}}, \ and\
  \bibinfo {author} {\bibfnamefont {L.}~\bibnamefont {Mandel}},\ }\href
  {\doibase 10.1103/PhysRevLett.59.2044} {\bibfield  {journal} {\bibinfo
  {journal} {Phys. Rev. Lett.}\ }\textbf {\bibinfo {volume} {59}},\ \bibinfo
  {pages} {2044} (\bibinfo {year} {1987})}\BibitemShut {NoStop}%
\bibitem [{\citenamefont {Tomonaga}(1950)}]{tomonaga}%
  \BibitemOpen
  \bibfield  {author} {\bibinfo {author} {\bibfnamefont {S.-i.}\ \bibnamefont
  {Tomonaga}},\ }\href {\doibase 10.1143/ptp/5.4.544} {\bibfield  {journal}
  {\bibinfo  {journal} {Progress of Theoretical Physics}\ }\textbf {\bibinfo
  {volume} {5}},\ \bibinfo {pages} {544} (\bibinfo {year} {1950})}\BibitemShut
  {NoStop}%
\bibitem [{\citenamefont {Luttinger}(1963)}]{luttinger}%
  \BibitemOpen
  \bibfield  {author} {\bibinfo {author} {\bibfnamefont {J.~M.}\ \bibnamefont
  {Luttinger}},\ }\href {\doibase 10.1063/1.1704046} {\bibfield  {journal}
  {\bibinfo  {journal} {Journal of Mathematical Physics}\ }\textbf {\bibinfo
  {volume} {4}},\ \bibinfo {pages} {1154} (\bibinfo {year} {1963})}\BibitemShut
  {NoStop}%
\bibitem [{\citenamefont {Haldane}(1981)}]{haldane81}%
  \BibitemOpen
  \bibfield  {author} {\bibinfo {author} {\bibfnamefont {F.~D.~M.}\
  \bibnamefont {Haldane}},\ }\href
  {http://stacks.iop.org/0022-3719/14/i=19/a=010} {\bibfield  {journal}
  {\bibinfo  {journal} {Journal of Physics C: Solid State Physics}\ }\textbf
  {\bibinfo {volume} {14}},\ \bibinfo {pages} {2585} (\bibinfo {year}
  {1981})}\BibitemShut {NoStop}%
\bibitem [{\citenamefont {Giamarchi}(2003)}]{giamarchi}%
  \BibitemOpen
  \bibfield  {author} {\bibinfo {author} {\bibfnamefont {T.}~\bibnamefont
  {Giamarchi}},\ }\href@noop {} {\emph {\bibinfo {title} {Quantum Physics in
  One Dimension}}}\ (\bibinfo  {publisher} {Oxford University Press},\ \bibinfo
  {year} {2003})\BibitemShut {NoStop}%
\bibitem [{\citenamefont {Auslaender}\ \emph {et~al.}(2002)\citenamefont
  {Auslaender}, \citenamefont {Yacoby}, \citenamefont {de~Picciotto},
  \citenamefont {Baldwin}, \citenamefont {Pfeiffer},\ and\ \citenamefont
  {West}}]{auslaender02}%
  \BibitemOpen
  \bibfield  {author} {\bibinfo {author} {\bibfnamefont {O.~M.}\ \bibnamefont
  {Auslaender}}, \bibinfo {author} {\bibfnamefont {A.}~\bibnamefont {Yacoby}},
  \bibinfo {author} {\bibfnamefont {R.}~\bibnamefont {de~Picciotto}}, \bibinfo
  {author} {\bibfnamefont {K.~W.}\ \bibnamefont {Baldwin}}, \bibinfo {author}
  {\bibfnamefont {L.~N.}\ \bibnamefont {Pfeiffer}}, \ and\ \bibinfo {author}
  {\bibfnamefont {K.~W.}\ \bibnamefont {West}},\ }\href {\doibase
  10.1126/science.1066266} {\bibfield  {journal} {\bibinfo  {journal}
  {Science}\ }\textbf {\bibinfo {volume} {295}},\ \bibinfo {pages} {825}
  (\bibinfo {year} {2002})}\BibitemShut {NoStop}%
\bibitem [{\citenamefont {Jompol}\ \emph {et~al.}(2009)\citenamefont {Jompol},
  \citenamefont {Ford}, \citenamefont {Griffiths}, \citenamefont {Farrer},
  \citenamefont {Jones}, \citenamefont {Anderson}, \citenamefont {Ritchie},
  \citenamefont {Silk},\ and\ \citenamefont {Schofield}}]{jompol-spincharge}%
  \BibitemOpen
  \bibfield  {author} {\bibinfo {author} {\bibfnamefont {Y.}~\bibnamefont
  {Jompol}}, \bibinfo {author} {\bibfnamefont {C.~J.~B.}\ \bibnamefont {Ford}},
  \bibinfo {author} {\bibfnamefont {J.~P.}\ \bibnamefont {Griffiths}}, \bibinfo
  {author} {\bibfnamefont {I.}~\bibnamefont {Farrer}}, \bibinfo {author}
  {\bibfnamefont {G.~A.~C.}\ \bibnamefont {Jones}}, \bibinfo {author}
  {\bibfnamefont {D.}~\bibnamefont {Anderson}}, \bibinfo {author}
  {\bibfnamefont {D.~A.}\ \bibnamefont {Ritchie}}, \bibinfo {author}
  {\bibfnamefont {T.~W.}\ \bibnamefont {Silk}}, \ and\ \bibinfo {author}
  {\bibfnamefont {A.~J.}\ \bibnamefont {Schofield}},\ }\href {\doibase
  10.1126/science.1171769} {\bibfield  {journal} {\bibinfo  {journal}
  {Science}\ }\textbf {\bibinfo {volume} {325}},\ \bibinfo {pages} {597}
  (\bibinfo {year} {2009})}\BibitemShut {NoStop}%
\bibitem [{\citenamefont {{Hashisaka}}\ \emph {et~al.}(2017)\citenamefont
  {{Hashisaka}}, \citenamefont {{Hiyama}}, \citenamefont {{Akiho}},
  \citenamefont {{Muraki}},\ and\ \citenamefont
  {{Fujisawa}}}]{hashisaka17spincharge}%
  \BibitemOpen
  \bibfield  {author} {\bibinfo {author} {\bibfnamefont {M.}~\bibnamefont
  {{Hashisaka}}}, \bibinfo {author} {\bibfnamefont {N.}~\bibnamefont
  {{Hiyama}}}, \bibinfo {author} {\bibfnamefont {T.}~\bibnamefont {{Akiho}}},
  \bibinfo {author} {\bibfnamefont {K.}~\bibnamefont {{Muraki}}}, \ and\
  \bibinfo {author} {\bibfnamefont {T.}~\bibnamefont {{Fujisawa}}},\ }\href
  {\doibase 10.1038/nphys4062} {\bibfield  {journal} {\bibinfo  {journal}
  {Nature Physics}\ }\textbf {\bibinfo {volume} {13}},\ \bibinfo {pages} {559}
  (\bibinfo {year} {2017})}\BibitemShut {NoStop}%
\bibitem [{\citenamefont {Maslov}\ and\ \citenamefont
  {Stone}(1995)}]{maslov-stone}%
  \BibitemOpen
  \bibfield  {author} {\bibinfo {author} {\bibfnamefont {D.~L.}\ \bibnamefont
  {Maslov}}\ and\ \bibinfo {author} {\bibfnamefont {M.}~\bibnamefont {Stone}},\
  }\href {\doibase 10.1103/PhysRevB.52.R5539} {\bibfield  {journal} {\bibinfo
  {journal} {Phys. Rev. B}\ }\textbf {\bibinfo {volume} {52}},\ \bibinfo
  {pages} {R5539} (\bibinfo {year} {1995})}\BibitemShut {NoStop}%
\bibitem [{\citenamefont {Safi}\ and\ \citenamefont {Schulz}(1995)}]{safi95}%
  \BibitemOpen
  \bibfield  {author} {\bibinfo {author} {\bibfnamefont {I.}~\bibnamefont
  {Safi}}\ and\ \bibinfo {author} {\bibfnamefont {H.~J.}\ \bibnamefont
  {Schulz}},\ }\href {\doibase 10.1103/PhysRevB.52.R17040} {\bibfield
  {journal} {\bibinfo  {journal} {Phys. Rev. B}\ }\textbf {\bibinfo {volume}
  {52}},\ \bibinfo {pages} {R17040} (\bibinfo {year} {1995})}\BibitemShut
  {NoStop}%
\bibitem [{\citenamefont {Safi}(1997)}]{safi97}%
  \BibitemOpen
  \bibfield  {author} {\bibinfo {author} {\bibfnamefont {I.}~\bibnamefont
  {Safi}},\ }\href {\doibase 10.1051/anphys:199705001} {\bibfield  {journal}
  {\bibinfo  {journal} {Ann. Phys. (France)}\ }\textbf {\bibinfo {volume}
  {22}},\ \bibinfo {pages} {463} (\bibinfo {year} {1997})}\BibitemShut
  {NoStop}%
\bibitem [{\citenamefont {Pham}\ \emph {et~al.}(2000)\citenamefont {Pham},
  \citenamefont {Gabay},\ and\ \citenamefont {Lederer}}]{pham2000}%
  \BibitemOpen
  \bibfield  {author} {\bibinfo {author} {\bibfnamefont {K.-V.}\ \bibnamefont
  {Pham}}, \bibinfo {author} {\bibfnamefont {M.}~\bibnamefont {Gabay}}, \ and\
  \bibinfo {author} {\bibfnamefont {P.}~\bibnamefont {Lederer}},\ }\href
  {\doibase 10.1103/PhysRevB.61.16397} {\bibfield  {journal} {\bibinfo
  {journal} {Phys. Rev. B}\ }\textbf {\bibinfo {volume} {61}},\ \bibinfo
  {pages} {16397} (\bibinfo {year} {2000})}\BibitemShut {NoStop}%
\bibitem [{\citenamefont {Steinberg}\ \emph {et~al.}(2008)\citenamefont
  {Steinberg}, \citenamefont {Barak}, \citenamefont {Yacoby}, \citenamefont
  {Pfeiffer}, \citenamefont {West}, \citenamefont {Halperin},\ and\
  \citenamefont {Le~Hur}}]{steinberg07chargefrac}%
  \BibitemOpen
  \bibfield  {author} {\bibinfo {author} {\bibfnamefont {H.}~\bibnamefont
  {Steinberg}}, \bibinfo {author} {\bibfnamefont {G.}~\bibnamefont {Barak}},
  \bibinfo {author} {\bibfnamefont {A.}~\bibnamefont {Yacoby}}, \bibinfo
  {author} {\bibfnamefont {L.~N.}\ \bibnamefont {Pfeiffer}}, \bibinfo {author}
  {\bibfnamefont {K.~W.}\ \bibnamefont {West}}, \bibinfo {author}
  {\bibfnamefont {B.~I.}\ \bibnamefont {Halperin}}, \ and\ \bibinfo {author}
  {\bibfnamefont {K.}~\bibnamefont {Le~Hur}},\ }\href {\doibase
  10.1038/nphys810} {\bibfield  {journal} {\bibinfo  {journal} {Nature
  Physics}\ }\textbf {\bibinfo {volume} {4}},\ \bibinfo {pages} {116} (\bibinfo
  {year} {2008})}\BibitemShut {NoStop}%
\bibitem [{\citenamefont {Kamata}\ \emph {et~al.}(2014)\citenamefont {Kamata},
  \citenamefont {Kumada}, \citenamefont {Hashisaka}, \citenamefont {Muraki},\
  and\ \citenamefont {Fujisawa}}]{kamata14}%
  \BibitemOpen
  \bibfield  {author} {\bibinfo {author} {\bibfnamefont {H.}~\bibnamefont
  {Kamata}}, \bibinfo {author} {\bibfnamefont {N.}~\bibnamefont {Kumada}},
  \bibinfo {author} {\bibfnamefont {M.}~\bibnamefont {Hashisaka}}, \bibinfo
  {author} {\bibfnamefont {K.}~\bibnamefont {Muraki}}, \ and\ \bibinfo {author}
  {\bibfnamefont {T.}~\bibnamefont {Fujisawa}},\ }\href {\doibase
  10.1038/nnano.2013.312} {\bibfield  {journal} {\bibinfo  {journal} {Nature
  Nanotechnology}\ }\textbf {\bibinfo {volume} {9}},\ \bibinfo {pages} {177}
  (\bibinfo {year} {2014})}\BibitemShut {NoStop}%
\bibitem [{\citenamefont {Perfetto}\ \emph {et~al.}(2014)\citenamefont
  {Perfetto}, \citenamefont {Stefanucci}, \citenamefont {Kamata},\ and\
  \citenamefont {Fujisawa}}]{perfetto14timeresolved}%
  \BibitemOpen
  \bibfield  {author} {\bibinfo {author} {\bibfnamefont {E.}~\bibnamefont
  {Perfetto}}, \bibinfo {author} {\bibfnamefont {G.}~\bibnamefont
  {Stefanucci}}, \bibinfo {author} {\bibfnamefont {H.}~\bibnamefont {Kamata}},
  \ and\ \bibinfo {author} {\bibfnamefont {T.}~\bibnamefont {Fujisawa}},\
  }\href {\doibase 10.1103/PhysRevB.89.201413} {\bibfield  {journal} {\bibinfo
  {journal} {Phys. Rev. B}\ }\textbf {\bibinfo {volume} {89}},\ \bibinfo
  {pages} {201413} (\bibinfo {year} {2014})}\BibitemShut {NoStop}%
\bibitem [{\citenamefont {{Calzona}}\ \emph {et~al.}(2015)\citenamefont
  {{Calzona}}, \citenamefont {{Carrega}}, \citenamefont {{Dolcetto}},\ and\
  \citenamefont {{Sassetti}}}]{calzona15physicaE}%
  \BibitemOpen
  \bibfield  {author} {\bibinfo {author} {\bibfnamefont {A.}~\bibnamefont
  {{Calzona}}}, \bibinfo {author} {\bibfnamefont {M.}~\bibnamefont
  {{Carrega}}}, \bibinfo {author} {\bibfnamefont {G.}~\bibnamefont
  {{Dolcetto}}}, \ and\ \bibinfo {author} {\bibfnamefont {M.}~\bibnamefont
  {{Sassetti}}},\ }\href {\doibase 10.1016/j.physe.2015.08.033} {\bibfield
  {journal} {\bibinfo  {journal} {Physica E Low-Dimensional Systems and
  Nanostructures}\ }\textbf {\bibinfo {volume} {74}},\ \bibinfo {pages} {630}
  (\bibinfo {year} {2015})}\BibitemShut {NoStop}%
\bibitem [{\citenamefont {Calzona}\ \emph {et~al.}(2015)\citenamefont
  {Calzona}, \citenamefont {Carrega}, \citenamefont {Dolcetto},\ and\
  \citenamefont {Sassetti}}]{calzona15spin}%
  \BibitemOpen
  \bibfield  {author} {\bibinfo {author} {\bibfnamefont {A.}~\bibnamefont
  {Calzona}}, \bibinfo {author} {\bibfnamefont {M.}~\bibnamefont {Carrega}},
  \bibinfo {author} {\bibfnamefont {G.}~\bibnamefont {Dolcetto}}, \ and\
  \bibinfo {author} {\bibfnamefont {M.}~\bibnamefont {Sassetti}},\ }\href
  {\doibase 10.1103/PhysRevB.92.195414} {\bibfield  {journal} {\bibinfo
  {journal} {Phys. Rev. B}\ }\textbf {\bibinfo {volume} {92}},\ \bibinfo
  {pages} {195414} (\bibinfo {year} {2015})}\BibitemShut {NoStop}%
\bibitem [{\citenamefont {Karzig}\ \emph {et~al.}(2011)\citenamefont {Karzig},
  \citenamefont {Refael}, \citenamefont {Glazman},\ and\ \citenamefont {von
  Oppen}}]{karzig11}%
  \BibitemOpen
  \bibfield  {author} {\bibinfo {author} {\bibfnamefont {T.}~\bibnamefont
  {Karzig}}, \bibinfo {author} {\bibfnamefont {G.}~\bibnamefont {Refael}},
  \bibinfo {author} {\bibfnamefont {L.~I.}\ \bibnamefont {Glazman}}, \ and\
  \bibinfo {author} {\bibfnamefont {F.}~\bibnamefont {von Oppen}},\ }\href
  {\doibase 10.1103/PhysRevLett.107.176403} {\bibfield  {journal} {\bibinfo
  {journal} {Phys. Rev. Lett.}\ }\textbf {\bibinfo {volume} {107}},\ \bibinfo
  {pages} {176403} (\bibinfo {year} {2011})}\BibitemShut {NoStop}%
\bibitem [{\citenamefont {Calzona}\ \emph {et~al.}(2017)\citenamefont
  {Calzona}, \citenamefont {Gambetta}, \citenamefont {Carrega}, \citenamefont
  {Cavaliere},\ and\ \citenamefont {Sassetti}}]{calzona17quench}%
  \BibitemOpen
  \bibfield  {author} {\bibinfo {author} {\bibfnamefont {A.}~\bibnamefont
  {Calzona}}, \bibinfo {author} {\bibfnamefont {F.~M.}\ \bibnamefont
  {Gambetta}}, \bibinfo {author} {\bibfnamefont {M.}~\bibnamefont {Carrega}},
  \bibinfo {author} {\bibfnamefont {F.}~\bibnamefont {Cavaliere}}, \ and\
  \bibinfo {author} {\bibfnamefont {M.}~\bibnamefont {Sassetti}},\ }\href
  {\doibase 10.1103/PhysRevB.95.085101} {\bibfield  {journal} {\bibinfo
  {journal} {Phys. Rev. B}\ }\textbf {\bibinfo {volume} {95}},\ \bibinfo
  {pages} {085101} (\bibinfo {year} {2017})}\BibitemShut {NoStop}%
\bibitem [{\citenamefont {Chalker}\ \emph {et~al.}(2007)\citenamefont
  {Chalker}, \citenamefont {Gefen},\ and\ \citenamefont
  {Veillette}}]{chalker2007MZ}%
  \BibitemOpen
  \bibfield  {author} {\bibinfo {author} {\bibfnamefont {J.~T.}\ \bibnamefont
  {Chalker}}, \bibinfo {author} {\bibfnamefont {Y.}~\bibnamefont {Gefen}}, \
  and\ \bibinfo {author} {\bibfnamefont {M.~Y.}\ \bibnamefont {Veillette}},\
  }\href {\doibase 10.1103/PhysRevB.76.085320} {\bibfield  {journal} {\bibinfo
  {journal} {Phys. Rev. B}\ }\textbf {\bibinfo {volume} {76}},\ \bibinfo
  {pages} {085320} (\bibinfo {year} {2007})}\BibitemShut {NoStop}%
\bibitem [{\citenamefont {Levkivskyi}\ and\ \citenamefont
  {Sukhorukov}(2008)}]{levkivskyi08modelnu2}%
  \BibitemOpen
  \bibfield  {author} {\bibinfo {author} {\bibfnamefont {I.~P.}\ \bibnamefont
  {Levkivskyi}}\ and\ \bibinfo {author} {\bibfnamefont {E.~V.}\ \bibnamefont
  {Sukhorukov}},\ }\href {\doibase 10.1103/PhysRevB.78.045322} {\bibfield
  {journal} {\bibinfo  {journal} {Phys. Rev. B}\ }\textbf {\bibinfo {volume}
  {78}},\ \bibinfo {pages} {045322} (\bibinfo {year} {2008})}\BibitemShut
  {NoStop}%
\bibitem [{\citenamefont {Kovrizhin}\ and\ \citenamefont
  {Chalker}(2009)}]{kovrizhin2009MZ}%
  \BibitemOpen
  \bibfield  {author} {\bibinfo {author} {\bibfnamefont {D.~L.}\ \bibnamefont
  {Kovrizhin}}\ and\ \bibinfo {author} {\bibfnamefont {J.~T.}\ \bibnamefont
  {Chalker}},\ }\href {\doibase 10.1103/PhysRevB.80.161306} {\bibfield
  {journal} {\bibinfo  {journal} {Phys. Rev. B}\ }\textbf {\bibinfo {volume}
  {80}},\ \bibinfo {pages} {161306} (\bibinfo {year} {2009})}\BibitemShut
  {NoStop}%
\bibitem [{\citenamefont {Kovrizhin}\ and\ \citenamefont
  {Chalker}(2010)}]{kovrizhin2010MZ}%
  \BibitemOpen
  \bibfield  {author} {\bibinfo {author} {\bibfnamefont {D.~L.}\ \bibnamefont
  {Kovrizhin}}\ and\ \bibinfo {author} {\bibfnamefont {J.~T.}\ \bibnamefont
  {Chalker}},\ }\href {\doibase 10.1103/PhysRevB.81.155318} {\bibfield
  {journal} {\bibinfo  {journal} {Phys. Rev. B}\ }\textbf {\bibinfo {volume}
  {81}},\ \bibinfo {pages} {155318} (\bibinfo {year} {2010})}\BibitemShut
  {NoStop}%
\bibitem [{\citenamefont {Degiovanni}\ \emph {et~al.}(2009)\citenamefont
  {Degiovanni}, \citenamefont {Grenier},\ and\ \citenamefont
  {F\`eve}}]{degiovanni2009relaxation}%
  \BibitemOpen
  \bibfield  {author} {\bibinfo {author} {\bibfnamefont {P.}~\bibnamefont
  {Degiovanni}}, \bibinfo {author} {\bibfnamefont {C.}~\bibnamefont {Grenier}},
  \ and\ \bibinfo {author} {\bibfnamefont {G.}~\bibnamefont {F\`eve}},\ }\href
  {\doibase 10.1103/PhysRevB.80.241307} {\bibfield  {journal} {\bibinfo
  {journal} {Phys. Rev. B}\ }\textbf {\bibinfo {volume} {80}},\ \bibinfo
  {pages} {241307} (\bibinfo {year} {2009})}\BibitemShut {NoStop}%
\bibitem [{\citenamefont {Degiovanni}\ \emph {et~al.}(2010)\citenamefont
  {Degiovanni}, \citenamefont {Grenier}, \citenamefont {F\`eve}, \citenamefont
  {Altimiras}, \citenamefont {le~Sueur},\ and\ \citenamefont
  {Pierre}}]{degiovanni2010relaxation}%
  \BibitemOpen
  \bibfield  {author} {\bibinfo {author} {\bibfnamefont {P.}~\bibnamefont
  {Degiovanni}}, \bibinfo {author} {\bibfnamefont {C.}~\bibnamefont {Grenier}},
  \bibinfo {author} {\bibfnamefont {G.}~\bibnamefont {F\`eve}}, \bibinfo
  {author} {\bibfnamefont {C.}~\bibnamefont {Altimiras}}, \bibinfo {author}
  {\bibfnamefont {H.}~\bibnamefont {le~Sueur}}, \ and\ \bibinfo {author}
  {\bibfnamefont {F.}~\bibnamefont {Pierre}},\ }\href {\doibase
  10.1103/PhysRevB.81.121302} {\bibfield  {journal} {\bibinfo  {journal} {Phys.
  Rev. B}\ }\textbf {\bibinfo {volume} {81}},\ \bibinfo {pages} {121302}
  (\bibinfo {year} {2010})}\BibitemShut {NoStop}%
\bibitem [{\citenamefont {Jonckheere}\ \emph {et~al.}(2012)\citenamefont
  {Jonckheere}, \citenamefont {Rech}, \citenamefont {Wahl},\ and\ \citenamefont
  {Martin}}]{jonckheere12hom}%
  \BibitemOpen
  \bibfield  {author} {\bibinfo {author} {\bibfnamefont {T.}~\bibnamefont
  {Jonckheere}}, \bibinfo {author} {\bibfnamefont {J.}~\bibnamefont {Rech}},
  \bibinfo {author} {\bibfnamefont {C.}~\bibnamefont {Wahl}}, \ and\ \bibinfo
  {author} {\bibfnamefont {T.}~\bibnamefont {Martin}},\ }\href {\doibase
  10.1103/PhysRevB.86.125425} {\bibfield  {journal} {\bibinfo  {journal} {Phys.
  Rev. B}\ }\textbf {\bibinfo {volume} {86}},\ \bibinfo {pages} {125425}
  (\bibinfo {year} {2012})}\BibitemShut {NoStop}%
\bibitem [{\citenamefont {Grenier}\ \emph {et~al.}(2013)\citenamefont
  {Grenier}, \citenamefont {Dubois}, \citenamefont {Jullien}, \citenamefont
  {Roulleau}, \citenamefont {Glattli},\ and\ \citenamefont
  {Degiovanni}}]{grenier13}%
  \BibitemOpen
  \bibfield  {author} {\bibinfo {author} {\bibfnamefont {C.}~\bibnamefont
  {Grenier}}, \bibinfo {author} {\bibfnamefont {J.}~\bibnamefont {Dubois}},
  \bibinfo {author} {\bibfnamefont {T.}~\bibnamefont {Jullien}}, \bibinfo
  {author} {\bibfnamefont {P.}~\bibnamefont {Roulleau}}, \bibinfo {author}
  {\bibfnamefont {D.~C.}\ \bibnamefont {Glattli}}, \ and\ \bibinfo {author}
  {\bibfnamefont {P.}~\bibnamefont {Degiovanni}},\ }\href {\doibase
  10.1103/PhysRevB.88.085302} {\bibfield  {journal} {\bibinfo  {journal} {Phys.
  Rev. B}\ }\textbf {\bibinfo {volume} {88}},\ \bibinfo {pages} {085302}
  (\bibinfo {year} {2013})}\BibitemShut {NoStop}%
\bibitem [{\citenamefont {Ferraro}\ \emph
  {et~al.}(2014{\natexlab{b}})\citenamefont {Ferraro}, \citenamefont {Roussel},
  \citenamefont {Cabart}, \citenamefont {Thibierge}, \citenamefont {F\`eve},
  \citenamefont {Grenier},\ and\ \citenamefont
  {Degiovanni}}]{ferraro14decoherence}%
  \BibitemOpen
  \bibfield  {author} {\bibinfo {author} {\bibfnamefont {D.}~\bibnamefont
  {Ferraro}}, \bibinfo {author} {\bibfnamefont {B.}~\bibnamefont {Roussel}},
  \bibinfo {author} {\bibfnamefont {C.}~\bibnamefont {Cabart}}, \bibinfo
  {author} {\bibfnamefont {E.}~\bibnamefont {Thibierge}}, \bibinfo {author}
  {\bibfnamefont {G.}~\bibnamefont {F\`eve}}, \bibinfo {author} {\bibfnamefont
  {C.}~\bibnamefont {Grenier}}, \ and\ \bibinfo {author} {\bibfnamefont
  {P.}~\bibnamefont {Degiovanni}},\ }\href {\doibase
  10.1103/PhysRevLett.113.166403} {\bibfield  {journal} {\bibinfo  {journal}
  {Phys. Rev. Lett.}\ }\textbf {\bibinfo {volume} {113}},\ \bibinfo {pages}
  {166403} (\bibinfo {year} {2014}{\natexlab{b}})}\BibitemShut {NoStop}%
\bibitem [{\citenamefont {Wahl}\ \emph {et~al.}(2014)\citenamefont {Wahl},
  \citenamefont {Rech}, \citenamefont {Jonckheere},\ and\ \citenamefont
  {Martin}}]{wahl14prl}%
  \BibitemOpen
  \bibfield  {author} {\bibinfo {author} {\bibfnamefont {C.}~\bibnamefont
  {Wahl}}, \bibinfo {author} {\bibfnamefont {J.}~\bibnamefont {Rech}}, \bibinfo
  {author} {\bibfnamefont {T.}~\bibnamefont {Jonckheere}}, \ and\ \bibinfo
  {author} {\bibfnamefont {T.}~\bibnamefont {Martin}},\ }\href {\doibase
  10.1103/PhysRevLett.112.046802} {\bibfield  {journal} {\bibinfo  {journal}
  {Phys. Rev. Lett.}\ }\textbf {\bibinfo {volume} {112}},\ \bibinfo {pages}
  {046802} (\bibinfo {year} {2014})}\BibitemShut {NoStop}%
\bibitem [{\citenamefont {Marguerite}\ \emph {et~al.}(2016)\citenamefont
  {Marguerite}, \citenamefont {Cabart}, \citenamefont {Wahl}, \citenamefont
  {Roussel}, \citenamefont {Freulon}, \citenamefont {Ferraro}, \citenamefont
  {Grenier}, \citenamefont {Berroir}, \citenamefont
  {Pla\ifmmode~\mbox{\c{c}}\else \c{c}\fi{}ais}, \citenamefont {Jonckheere},
  \citenamefont {Rech}, \citenamefont {Martin}, \citenamefont {Degiovanni},
  \citenamefont {Cavanna}, \citenamefont {Jin},\ and\ \citenamefont
  {F\`eve}}]{marguerite16decoherence}%
  \BibitemOpen
  \bibfield  {author} {\bibinfo {author} {\bibfnamefont {A.}~\bibnamefont
  {Marguerite}}, \bibinfo {author} {\bibfnamefont {C.}~\bibnamefont {Cabart}},
  \bibinfo {author} {\bibfnamefont {C.}~\bibnamefont {Wahl}}, \bibinfo {author}
  {\bibfnamefont {B.}~\bibnamefont {Roussel}}, \bibinfo {author} {\bibfnamefont
  {V.}~\bibnamefont {Freulon}}, \bibinfo {author} {\bibfnamefont
  {D.}~\bibnamefont {Ferraro}}, \bibinfo {author} {\bibfnamefont
  {C.}~\bibnamefont {Grenier}}, \bibinfo {author} {\bibfnamefont {J.-M.}\
  \bibnamefont {Berroir}}, \bibinfo {author} {\bibfnamefont {B.}~\bibnamefont
  {Pla\ifmmode~\mbox{\c{c}}\else \c{c}\fi{}ais}}, \bibinfo {author}
  {\bibfnamefont {T.}~\bibnamefont {Jonckheere}}, \bibinfo {author}
  {\bibfnamefont {J.}~\bibnamefont {Rech}}, \bibinfo {author} {\bibfnamefont
  {T.}~\bibnamefont {Martin}}, \bibinfo {author} {\bibfnamefont
  {P.}~\bibnamefont {Degiovanni}}, \bibinfo {author} {\bibfnamefont
  {A.}~\bibnamefont {Cavanna}}, \bibinfo {author} {\bibfnamefont
  {Y.}~\bibnamefont {Jin}}, \ and\ \bibinfo {author} {\bibfnamefont
  {G.}~\bibnamefont {F\`eve}},\ }\href {\doibase 10.1103/PhysRevB.94.115311}
  {\bibfield  {journal} {\bibinfo  {journal} {Phys. Rev. B}\ }\textbf {\bibinfo
  {volume} {94}},\ \bibinfo {pages} {115311} (\bibinfo {year}
  {2016})}\BibitemShut {NoStop}%
\bibitem [{\citenamefont {Slobodeniuk}\ \emph {et~al.}(2016)\citenamefont
  {Slobodeniuk}, \citenamefont {Idrisov},\ and\ \citenamefont
  {Sukhorukov}}]{sukhorukov2016prb}%
  \BibitemOpen
  \bibfield  {author} {\bibinfo {author} {\bibfnamefont {A.~O.}\ \bibnamefont
  {Slobodeniuk}}, \bibinfo {author} {\bibfnamefont {E.~G.}\ \bibnamefont
  {Idrisov}}, \ and\ \bibinfo {author} {\bibfnamefont {E.~V.}\ \bibnamefont
  {Sukhorukov}},\ }\href {\doibase 10.1103/PhysRevB.93.035421} {\bibfield
  {journal} {\bibinfo  {journal} {Phys. Rev. B}\ }\textbf {\bibinfo {volume}
  {93}},\ \bibinfo {pages} {035421} (\bibinfo {year} {2016})}\BibitemShut
  {NoStop}%
\bibitem [{\citenamefont {{Cabart}}\ \emph {et~al.}()\citenamefont {{Cabart}},
  \citenamefont {{Roussel}}, \citenamefont {{F{\`e}ve}},\ and\ \citenamefont
  {{Degiovanni}}}]{cabart18}%
  \BibitemOpen
  \bibfield  {author} {\bibinfo {author} {\bibfnamefont {C.}~\bibnamefont
  {{Cabart}}}, \bibinfo {author} {\bibfnamefont {B.}~\bibnamefont {{Roussel}}},
  \bibinfo {author} {\bibfnamefont {G.}~\bibnamefont {{F{\`e}ve}}}, \ and\
  \bibinfo {author} {\bibfnamefont {P.}~\bibnamefont {{Degiovanni}}},\
  }\href@noop {} {\ }\Eprint {http://arxiv.org/abs/1804.04054}
  {arXiv:1804.04054} \BibitemShut {NoStop}%
\bibitem [{\citenamefont {Kovrizhin}\ and\ \citenamefont
  {Chalker}(2011)}]{kovrizhin2011}%
  \BibitemOpen
  \bibfield  {author} {\bibinfo {author} {\bibfnamefont {D.~L.}\ \bibnamefont
  {Kovrizhin}}\ and\ \bibinfo {author} {\bibfnamefont {J.~T.}\ \bibnamefont
  {Chalker}},\ }\href {\doibase 10.1103/PhysRevB.84.085105} {\bibfield
  {journal} {\bibinfo  {journal} {Phys. Rev. B}\ }\textbf {\bibinfo {volume}
  {84}},\ \bibinfo {pages} {085105} (\bibinfo {year} {2011})}\BibitemShut
  {NoStop}%
\bibitem [{\citenamefont {Kovrizhin}\ and\ \citenamefont
  {Chalker}(2012)}]{kovrizhin2012}%
  \BibitemOpen
  \bibfield  {author} {\bibinfo {author} {\bibfnamefont {D.~L.}\ \bibnamefont
  {Kovrizhin}}\ and\ \bibinfo {author} {\bibfnamefont {J.~T.}\ \bibnamefont
  {Chalker}},\ }\href {\doibase 10.1103/PhysRevLett.109.106403} {\bibfield
  {journal} {\bibinfo  {journal} {Phys. Rev. Lett.}\ }\textbf {\bibinfo
  {volume} {109}},\ \bibinfo {pages} {106403} (\bibinfo {year}
  {2012})}\BibitemShut {NoStop}%
\bibitem [{\citenamefont {Levkivskyi}\ and\ \citenamefont
  {Sukhorukov}(2012)}]{levkivskyi12relaxation}%
  \BibitemOpen
  \bibfield  {author} {\bibinfo {author} {\bibfnamefont {I.~P.}\ \bibnamefont
  {Levkivskyi}}\ and\ \bibinfo {author} {\bibfnamefont {E.~V.}\ \bibnamefont
  {Sukhorukov}},\ }\href {\doibase 10.1103/PhysRevB.85.075309} {\bibfield
  {journal} {\bibinfo  {journal} {Phys. Rev. B}\ }\textbf {\bibinfo {volume}
  {85}},\ \bibinfo {pages} {075309} (\bibinfo {year} {2012})}\BibitemShut
  {NoStop}%
\bibitem [{\citenamefont {le~Sueur}\ \emph {et~al.}(2010)\citenamefont
  {le~Sueur}, \citenamefont {Altimiras}, \citenamefont {Gennser}, \citenamefont
  {Cavanna}, \citenamefont {Mailly},\ and\ \citenamefont
  {Pierre}}]{sueur2010relaxation}%
  \BibitemOpen
  \bibfield  {author} {\bibinfo {author} {\bibfnamefont {H.}~\bibnamefont
  {le~Sueur}}, \bibinfo {author} {\bibfnamefont {C.}~\bibnamefont {Altimiras}},
  \bibinfo {author} {\bibfnamefont {U.}~\bibnamefont {Gennser}}, \bibinfo
  {author} {\bibfnamefont {A.}~\bibnamefont {Cavanna}}, \bibinfo {author}
  {\bibfnamefont {D.}~\bibnamefont {Mailly}}, \ and\ \bibinfo {author}
  {\bibfnamefont {F.}~\bibnamefont {Pierre}},\ }\href {\doibase
  10.1103/PhysRevLett.105.056803} {\bibfield  {journal} {\bibinfo  {journal}
  {Phys. Rev. Lett.}\ }\textbf {\bibinfo {volume} {105}},\ \bibinfo {pages}
  {056803} (\bibinfo {year} {2010})}\BibitemShut {NoStop}%
\bibitem [{\citenamefont {{Altimiras}}\ \emph {et~al.}(2010)\citenamefont
  {{Altimiras}}, \citenamefont {{Le Sueur}}, \citenamefont {{Gennser}},
  \citenamefont {{Cavanna}}, \citenamefont {{Mailly}},\ and\ \citenamefont
  {{Pierre}}}]{altimiras2010relaxation}%
  \BibitemOpen
  \bibfield  {author} {\bibinfo {author} {\bibfnamefont {C.}~\bibnamefont
  {{Altimiras}}}, \bibinfo {author} {\bibfnamefont {H.}~\bibnamefont {{Le
  Sueur}}}, \bibinfo {author} {\bibfnamefont {U.}~\bibnamefont {{Gennser}}},
  \bibinfo {author} {\bibfnamefont {A.}~\bibnamefont {{Cavanna}}}, \bibinfo
  {author} {\bibfnamefont {D.}~\bibnamefont {{Mailly}}}, \ and\ \bibinfo
  {author} {\bibfnamefont {F.}~\bibnamefont {{Pierre}}},\ }\href {\doibase
  10.1038/nphys1429} {\bibfield  {journal} {\bibinfo  {journal} {Nature
  Physics}\ }\textbf {\bibinfo {volume} {6}},\ \bibinfo {pages} {34} (\bibinfo
  {year} {2010})}\BibitemShut {NoStop}%
\bibitem [{\citenamefont {Altimiras}\ \emph {et~al.}(2010)\citenamefont
  {Altimiras}, \citenamefont {le~Sueur}, \citenamefont {Gennser}, \citenamefont
  {Cavanna}, \citenamefont {Mailly},\ and\ \citenamefont
  {Pierre}}]{altimiras2010relaxtuning}%
  \BibitemOpen
  \bibfield  {author} {\bibinfo {author} {\bibfnamefont {C.}~\bibnamefont
  {Altimiras}}, \bibinfo {author} {\bibfnamefont {H.}~\bibnamefont {le~Sueur}},
  \bibinfo {author} {\bibfnamefont {U.}~\bibnamefont {Gennser}}, \bibinfo
  {author} {\bibfnamefont {A.}~\bibnamefont {Cavanna}}, \bibinfo {author}
  {\bibfnamefont {D.}~\bibnamefont {Mailly}}, \ and\ \bibinfo {author}
  {\bibfnamefont {F.}~\bibnamefont {Pierre}},\ }\href {\doibase
  10.1103/PhysRevLett.105.226804} {\bibfield  {journal} {\bibinfo  {journal}
  {Phys. Rev. Lett.}\ }\textbf {\bibinfo {volume} {105}},\ \bibinfo {pages}
  {226804} (\bibinfo {year} {2010})}\BibitemShut {NoStop}%
\bibitem [{\citenamefont {Itoh}\ \emph {et~al.}(2018)\citenamefont {Itoh},
  \citenamefont {Nakazawa}, \citenamefont {Ota}, \citenamefont {Hashisaka},
  \citenamefont {Muraki},\ and\ \citenamefont
  {Fujisawa}}]{itoh18-metastable-Hall}%
  \BibitemOpen
  \bibfield  {author} {\bibinfo {author} {\bibfnamefont {K.}~\bibnamefont
  {Itoh}}, \bibinfo {author} {\bibfnamefont {R.}~\bibnamefont {Nakazawa}},
  \bibinfo {author} {\bibfnamefont {T.}~\bibnamefont {Ota}}, \bibinfo {author}
  {\bibfnamefont {M.}~\bibnamefont {Hashisaka}}, \bibinfo {author}
  {\bibfnamefont {K.}~\bibnamefont {Muraki}}, \ and\ \bibinfo {author}
  {\bibfnamefont {T.}~\bibnamefont {Fujisawa}},\ }\href {\doibase
  10.1103/PhysRevLett.120.197701} {\bibfield  {journal} {\bibinfo  {journal}
  {Phys. Rev. Lett.}\ }\textbf {\bibinfo {volume} {120}},\ \bibinfo {pages}
  {197701} (\bibinfo {year} {2018})}\BibitemShut {NoStop}%
\bibitem [{\citenamefont {Paradiso}\ \emph {et~al.}(2011)\citenamefont
  {Paradiso}, \citenamefont {Heun}, \citenamefont {Roddaro}, \citenamefont
  {Sorba}, \citenamefont {Beltram},\ and\ \citenamefont
  {Biasiol}}]{paradiso2011}%
  \BibitemOpen
  \bibfield  {author} {\bibinfo {author} {\bibfnamefont {N.}~\bibnamefont
  {Paradiso}}, \bibinfo {author} {\bibfnamefont {S.}~\bibnamefont {Heun}},
  \bibinfo {author} {\bibfnamefont {S.}~\bibnamefont {Roddaro}}, \bibinfo
  {author} {\bibfnamefont {L.}~\bibnamefont {Sorba}}, \bibinfo {author}
  {\bibfnamefont {F.}~\bibnamefont {Beltram}}, \ and\ \bibinfo {author}
  {\bibfnamefont {G.}~\bibnamefont {Biasiol}},\ }\href {\doibase
  10.1103/PhysRevB.84.235318} {\bibfield  {journal} {\bibinfo  {journal} {Phys.
  Rev. B}\ }\textbf {\bibinfo {volume} {84}},\ \bibinfo {pages} {235318}
  (\bibinfo {year} {2011})}\BibitemShut {NoStop}%
\bibitem [{\citenamefont {Paradiso}\ \emph {et~al.}(2012)\citenamefont
  {Paradiso}, \citenamefont {Heun}, \citenamefont {Roddaro}, \citenamefont
  {Sorba}, \citenamefont {Beltram}, \citenamefont {Biasiol}, \citenamefont
  {Pfeiffer},\ and\ \citenamefont {West}}]{paradiso2012prl}%
  \BibitemOpen
  \bibfield  {author} {\bibinfo {author} {\bibfnamefont {N.}~\bibnamefont
  {Paradiso}}, \bibinfo {author} {\bibfnamefont {S.}~\bibnamefont {Heun}},
  \bibinfo {author} {\bibfnamefont {S.}~\bibnamefont {Roddaro}}, \bibinfo
  {author} {\bibfnamefont {L.}~\bibnamefont {Sorba}}, \bibinfo {author}
  {\bibfnamefont {F.}~\bibnamefont {Beltram}}, \bibinfo {author} {\bibfnamefont
  {G.}~\bibnamefont {Biasiol}}, \bibinfo {author} {\bibfnamefont {L.~N.}\
  \bibnamefont {Pfeiffer}}, \ and\ \bibinfo {author} {\bibfnamefont {K.~W.}\
  \bibnamefont {West}},\ }\href {\doibase 10.1103/PhysRevLett.108.246801}
  {\bibfield  {journal} {\bibinfo  {journal} {Phys. Rev. Lett.}\ }\textbf
  {\bibinfo {volume} {108}},\ \bibinfo {pages} {246801} (\bibinfo {year}
  {2012})}\BibitemShut {NoStop}%
\bibitem [{\citenamefont {{Guiducci}}\ \emph {et~al.}()\citenamefont
  {{Guiducci}}, \citenamefont {{Carrega}}, \citenamefont {{Biasiol}},
  \citenamefont {{Sorba}}, \citenamefont {{Beltram}},\ and\ \citenamefont
  {{Heun}}}]{guiducci2018}%
  \BibitemOpen
  \bibfield  {author} {\bibinfo {author} {\bibfnamefont {S.}~\bibnamefont
  {{Guiducci}}}, \bibinfo {author} {\bibfnamefont {M.}~\bibnamefont
  {{Carrega}}}, \bibinfo {author} {\bibfnamefont {G.}~\bibnamefont
  {{Biasiol}}}, \bibinfo {author} {\bibfnamefont {L.}~\bibnamefont {{Sorba}}},
  \bibinfo {author} {\bibfnamefont {F.}~\bibnamefont {{Beltram}}}, \ and\
  \bibinfo {author} {\bibfnamefont {S.}~\bibnamefont {{Heun}}},\ }\href@noop {}
  {\ }\Eprint {http://arxiv.org/abs/1805.02862} {arXiv:1805.02862} \BibitemShut
  {NoStop}%
\bibitem [{\citenamefont {Berg}\ \emph {et~al.}(2009)\citenamefont {Berg},
  \citenamefont {Oreg}, \citenamefont {Kim},\ and\ \citenamefont {von
  Oppen}}]{berg2009fractionalization}%
  \BibitemOpen
  \bibfield  {author} {\bibinfo {author} {\bibfnamefont {E.}~\bibnamefont
  {Berg}}, \bibinfo {author} {\bibfnamefont {Y.}~\bibnamefont {Oreg}}, \bibinfo
  {author} {\bibfnamefont {E.-A.}\ \bibnamefont {Kim}}, \ and\ \bibinfo
  {author} {\bibfnamefont {F.}~\bibnamefont {von Oppen}},\ }\href {\doibase
  10.1103/PhysRevLett.102.236402} {\bibfield  {journal} {\bibinfo  {journal}
  {Phys. Rev. Lett.}\ }\textbf {\bibinfo {volume} {102}},\ \bibinfo {pages}
  {236402} (\bibinfo {year} {2009})}\BibitemShut {NoStop}%
\bibitem [{\citenamefont {Neder}(2012)}]{neder2012}%
  \BibitemOpen
  \bibfield  {author} {\bibinfo {author} {\bibfnamefont {I.}~\bibnamefont
  {Neder}},\ }\href {\doibase 10.1103/PhysRevLett.108.186404} {\bibfield
  {journal} {\bibinfo  {journal} {Phys. Rev. Lett.}\ }\textbf {\bibinfo
  {volume} {108}},\ \bibinfo {pages} {186404} (\bibinfo {year}
  {2012})}\BibitemShut {NoStop}%
\bibitem [{\citenamefont {Milletar\`{\i}}\ and\ \citenamefont
  {Rosenow}(2013)}]{milletari13}%
  \BibitemOpen
  \bibfield  {author} {\bibinfo {author} {\bibfnamefont {M.}~\bibnamefont
  {Milletar\`{\i}}}\ and\ \bibinfo {author} {\bibfnamefont {B.}~\bibnamefont
  {Rosenow}},\ }\href {\doibase 10.1103/PhysRevLett.111.136807} {\bibfield
  {journal} {\bibinfo  {journal} {Phys. Rev. Lett.}\ }\textbf {\bibinfo
  {volume} {111}},\ \bibinfo {pages} {136807} (\bibinfo {year}
  {2013})}\BibitemShut {NoStop}%
\bibitem [{\citenamefont {Bocquillon}\ \emph
  {et~al.}(2013{\natexlab{b}})\citenamefont {Bocquillon}, \citenamefont
  {Freulon}, \citenamefont {Berroir}, \citenamefont {Degiovanni}, \citenamefont
  {Pla\c{c}ais}, \citenamefont {Cavanna}, \citenamefont {Jin},\ and\
  \citenamefont {F\`eve}}]{bocquillon2013}%
  \BibitemOpen
  \bibfield  {author} {\bibinfo {author} {\bibfnamefont {E.}~\bibnamefont
  {Bocquillon}}, \bibinfo {author} {\bibfnamefont {V.}~\bibnamefont {Freulon}},
  \bibinfo {author} {\bibfnamefont {J.-M.}\ \bibnamefont {Berroir}}, \bibinfo
  {author} {\bibfnamefont {P.}~\bibnamefont {Degiovanni}}, \bibinfo {author}
  {\bibfnamefont {B.}~\bibnamefont {Pla\c{c}ais}}, \bibinfo {author}
  {\bibfnamefont {A.}~\bibnamefont {Cavanna}}, \bibinfo {author} {\bibfnamefont
  {Y.}~\bibnamefont {Jin}}, \ and\ \bibinfo {author} {\bibfnamefont
  {G.}~\bibnamefont {F\`eve}},\ }\href {\doibase doi:10.1038/ncomms2788}
  {\bibfield  {journal} {\bibinfo  {journal} {Nature Communications}\ }\textbf
  {\bibinfo {volume} {4}},\ \bibinfo {pages} {1839} (\bibinfo {year}
  {2013}{\natexlab{b}})}\BibitemShut {NoStop}%
\bibitem [{\citenamefont {Inoue}\ \emph {et~al.}(2014)\citenamefont {Inoue},
  \citenamefont {Grivnin}, \citenamefont {Ofek}, \citenamefont {Neder},
  \citenamefont {Heiblum}, \citenamefont {Umansky},\ and\ \citenamefont
  {Mahalu}}]{inoue2014}%
  \BibitemOpen
  \bibfield  {author} {\bibinfo {author} {\bibfnamefont {H.}~\bibnamefont
  {Inoue}}, \bibinfo {author} {\bibfnamefont {A.}~\bibnamefont {Grivnin}},
  \bibinfo {author} {\bibfnamefont {N.}~\bibnamefont {Ofek}}, \bibinfo {author}
  {\bibfnamefont {I.}~\bibnamefont {Neder}}, \bibinfo {author} {\bibfnamefont
  {M.}~\bibnamefont {Heiblum}}, \bibinfo {author} {\bibfnamefont
  {V.}~\bibnamefont {Umansky}}, \ and\ \bibinfo {author} {\bibfnamefont
  {D.}~\bibnamefont {Mahalu}},\ }\href {\doibase
  10.1103/PhysRevLett.112.166801} {\bibfield  {journal} {\bibinfo  {journal}
  {Phys. Rev. Lett.}\ }\textbf {\bibinfo {volume} {112}},\ \bibinfo {pages}
  {166801} (\bibinfo {year} {2014})}\BibitemShut {NoStop}%
\bibitem [{\citenamefont {von Delft}\ and\ \citenamefont
  {Schoeller}(1998)}]{vondelft}%
  \BibitemOpen
  \bibfield  {author} {\bibinfo {author} {\bibfnamefont {J.}~\bibnamefont {von
  Delft}}\ and\ \bibinfo {author} {\bibfnamefont {H.}~\bibnamefont
  {Schoeller}},\ }\href {\doibase
  10.1002/(SICI)1521-3889(199811)7:4<225::AID-ANDP225>3.0.CO;2-L} {\bibfield
  {journal} {\bibinfo  {journal} {Annalen der Physik}\ }\textbf {\bibinfo
  {volume} {7}},\ \bibinfo {pages} {225} (\bibinfo {year} {1998})}\BibitemShut
  {NoStop}%
\bibitem [{\citenamefont {Guyon}\ \emph {et~al.}(2002)\citenamefont {Guyon},
  \citenamefont {Devillard}, \citenamefont {Martin},\ and\ \citenamefont
  {Safi}}]{safi02klein}%
  \BibitemOpen
  \bibfield  {author} {\bibinfo {author} {\bibfnamefont {R.}~\bibnamefont
  {Guyon}}, \bibinfo {author} {\bibfnamefont {P.}~\bibnamefont {Devillard}},
  \bibinfo {author} {\bibfnamefont {T.}~\bibnamefont {Martin}}, \ and\ \bibinfo
  {author} {\bibfnamefont {I.}~\bibnamefont {Safi}},\ }\href {\doibase
  10.1103/PhysRevB.65.153304} {\bibfield  {journal} {\bibinfo  {journal} {Phys.
  Rev. B}\ }\textbf {\bibinfo {volume} {65}},\ \bibinfo {pages} {153304}
  (\bibinfo {year} {2002})}\BibitemShut {NoStop}%
\bibitem [{\citenamefont {Markhof}\ and\ \citenamefont
  {Meden}(2016)}]{meden2016}%
  \BibitemOpen
  \bibfield  {author} {\bibinfo {author} {\bibfnamefont {L.}~\bibnamefont
  {Markhof}}\ and\ \bibinfo {author} {\bibfnamefont {V.}~\bibnamefont
  {Meden}},\ }\href {\doibase 10.1103/PhysRevB.93.085108} {\bibfield  {journal}
  {\bibinfo  {journal} {Phys. Rev. B}\ }\textbf {\bibinfo {volume} {93}},\
  \bibinfo {pages} {085108} (\bibinfo {year} {2016})}\BibitemShut {NoStop}%
\bibitem [{\citenamefont {Dubois}\ \emph
  {et~al.}(2013{\natexlab{b}})\citenamefont {Dubois}, \citenamefont {Jullien},
  \citenamefont {Grenier}, \citenamefont {Degiovanni}, \citenamefont
  {Roulleau},\ and\ \citenamefont {Glattli}}]{dubois13prb}%
  \BibitemOpen
  \bibfield  {author} {\bibinfo {author} {\bibfnamefont {J.}~\bibnamefont
  {Dubois}}, \bibinfo {author} {\bibfnamefont {T.}~\bibnamefont {Jullien}},
  \bibinfo {author} {\bibfnamefont {C.}~\bibnamefont {Grenier}}, \bibinfo
  {author} {\bibfnamefont {P.}~\bibnamefont {Degiovanni}}, \bibinfo {author}
  {\bibfnamefont {P.}~\bibnamefont {Roulleau}}, \ and\ \bibinfo {author}
  {\bibfnamefont {D.~C.}\ \bibnamefont {Glattli}},\ }\href {\doibase
  10.1103/PhysRevB.88.085301} {\bibfield  {journal} {\bibinfo  {journal} {Phys.
  Rev. B}\ }\textbf {\bibinfo {volume} {88}},\ \bibinfo {pages} {085301}
  (\bibinfo {year} {2013}{\natexlab{b}})}\BibitemShut {NoStop}%
\bibitem [{\citenamefont {Rech}\ \emph {et~al.}(2017)\citenamefont {Rech},
  \citenamefont {Ferraro}, \citenamefont {Jonckheere}, \citenamefont
  {Vannucci}, \citenamefont {Sassetti},\ and\ \citenamefont
  {Martin}}]{rech16prl}%
  \BibitemOpen
  \bibfield  {author} {\bibinfo {author} {\bibfnamefont {J.}~\bibnamefont
  {Rech}}, \bibinfo {author} {\bibfnamefont {D.}~\bibnamefont {Ferraro}},
  \bibinfo {author} {\bibfnamefont {T.}~\bibnamefont {Jonckheere}}, \bibinfo
  {author} {\bibfnamefont {L.}~\bibnamefont {Vannucci}}, \bibinfo {author}
  {\bibfnamefont {M.}~\bibnamefont {Sassetti}}, \ and\ \bibinfo {author}
  {\bibfnamefont {T.}~\bibnamefont {Martin}},\ }\href {\doibase
  10.1103/PhysRevLett.118.076801} {\bibfield  {journal} {\bibinfo  {journal}
  {Phys. Rev. Lett.}\ }\textbf {\bibinfo {volume} {118}},\ \bibinfo {pages}
  {076801} (\bibinfo {year} {2017})}\BibitemShut {NoStop}%
\bibitem [{\citenamefont {Vannucci}\ \emph {et~al.}(2017)\citenamefont
  {Vannucci}, \citenamefont {Ronetti}, \citenamefont {Rech}, \citenamefont
  {Ferraro}, \citenamefont {Jonckheere}, \citenamefont {Martin},\ and\
  \citenamefont {Sassetti}}]{vannucci17heat}%
  \BibitemOpen
  \bibfield  {author} {\bibinfo {author} {\bibfnamefont {L.}~\bibnamefont
  {Vannucci}}, \bibinfo {author} {\bibfnamefont {F.}~\bibnamefont {Ronetti}},
  \bibinfo {author} {\bibfnamefont {J.}~\bibnamefont {Rech}}, \bibinfo {author}
  {\bibfnamefont {D.}~\bibnamefont {Ferraro}}, \bibinfo {author} {\bibfnamefont
  {T.}~\bibnamefont {Jonckheere}}, \bibinfo {author} {\bibfnamefont
  {T.}~\bibnamefont {Martin}}, \ and\ \bibinfo {author} {\bibfnamefont
  {M.}~\bibnamefont {Sassetti}},\ }\href {\doibase 10.1103/PhysRevB.95.245415}
  {\bibfield  {journal} {\bibinfo  {journal} {Phys. Rev. B}\ }\textbf {\bibinfo
  {volume} {95}},\ \bibinfo {pages} {245415} (\bibinfo {year}
  {2017})}\BibitemShut {NoStop}%
\bibitem [{\citenamefont {Ronetti}\ \emph {et~al.}(2017)\citenamefont
  {Ronetti}, \citenamefont {Carrega}, \citenamefont {Ferraro}, \citenamefont
  {Rech}, \citenamefont {Jonckheere}, \citenamefont {Martin},\ and\
  \citenamefont {Sassetti}}]{ronetti17polarized}%
  \BibitemOpen
  \bibfield  {author} {\bibinfo {author} {\bibfnamefont {F.}~\bibnamefont
  {Ronetti}}, \bibinfo {author} {\bibfnamefont {M.}~\bibnamefont {Carrega}},
  \bibinfo {author} {\bibfnamefont {D.}~\bibnamefont {Ferraro}}, \bibinfo
  {author} {\bibfnamefont {J.}~\bibnamefont {Rech}}, \bibinfo {author}
  {\bibfnamefont {T.}~\bibnamefont {Jonckheere}}, \bibinfo {author}
  {\bibfnamefont {T.}~\bibnamefont {Martin}}, \ and\ \bibinfo {author}
  {\bibfnamefont {M.}~\bibnamefont {Sassetti}},\ }\href {\doibase
  10.1103/PhysRevB.95.115412} {\bibfield  {journal} {\bibinfo  {journal} {Phys.
  Rev. B}\ }\textbf {\bibinfo {volume} {95}},\ \bibinfo {pages} {115412}
  (\bibinfo {year} {2017})}\BibitemShut {NoStop}%
\bibitem [{\citenamefont {Keeling}\ \emph {et~al.}(2006)\citenamefont
  {Keeling}, \citenamefont {Klich},\ and\ \citenamefont {Levitov}}]{keeling06}%
  \BibitemOpen
  \bibfield  {author} {\bibinfo {author} {\bibfnamefont {J.}~\bibnamefont
  {Keeling}}, \bibinfo {author} {\bibfnamefont {I.}~\bibnamefont {Klich}}, \
  and\ \bibinfo {author} {\bibfnamefont {L.~S.}\ \bibnamefont {Levitov}},\
  }\href {\doibase 10.1103/PhysRevLett.97.116403} {\bibfield  {journal}
  {\bibinfo  {journal} {Phys. Rev. Lett.}\ }\textbf {\bibinfo {volume} {97}},\
  \bibinfo {pages} {116403} (\bibinfo {year} {2006})}\BibitemShut {NoStop}%
\bibitem [{\citenamefont {{Ronetti}}\ \emph {et~al.}()\citenamefont
  {{Ronetti}}, \citenamefont {{Vannucci}}, \citenamefont {{Ferraro}},
  \citenamefont {{Jonckheere}}, \citenamefont {{Rech}}, \citenamefont
  {{Martin}},\ and\ \citenamefont {{Sassetti}}}]{ronetti17levitons}%
  \BibitemOpen
  \bibfield  {author} {\bibinfo {author} {\bibfnamefont {F.}~\bibnamefont
  {{Ronetti}}}, \bibinfo {author} {\bibfnamefont {L.}~\bibnamefont
  {{Vannucci}}}, \bibinfo {author} {\bibfnamefont {D.}~\bibnamefont
  {{Ferraro}}}, \bibinfo {author} {\bibfnamefont {T.}~\bibnamefont
  {{Jonckheere}}}, \bibinfo {author} {\bibfnamefont {J.}~\bibnamefont
  {{Rech}}}, \bibinfo {author} {\bibfnamefont {T.}~\bibnamefont {{Martin}}}, \
  and\ \bibinfo {author} {\bibfnamefont {M.}~\bibnamefont {{Sassetti}}},\
  }\href@noop {} {\ }\Eprint {http://arxiv.org/abs/1712.07094}
  {arXiv:1712.07094} \BibitemShut {NoStop}%
\bibitem [{\citenamefont {Ferraro}\ \emph
  {et~al.}(2014{\natexlab{c}})\citenamefont {Ferraro}, \citenamefont {Carrega},
  \citenamefont {Braggio},\ and\ \citenamefont {Sassetti}}]{ferraro2014noise}%
  \BibitemOpen
  \bibfield  {author} {\bibinfo {author} {\bibfnamefont {D.}~\bibnamefont
  {Ferraro}}, \bibinfo {author} {\bibfnamefont {M.}~\bibnamefont {Carrega}},
  \bibinfo {author} {\bibfnamefont {A.}~\bibnamefont {Braggio}}, \ and\
  \bibinfo {author} {\bibfnamefont {M.}~\bibnamefont {Sassetti}},\ }\href
  {http://stacks.iop.org/1367-2630/16/i=4/a=043018} {\bibfield  {journal}
  {\bibinfo  {journal} {New Journal of Physics}\ }\textbf {\bibinfo {volume}
  {16}},\ \bibinfo {pages} {043018} (\bibinfo {year}
  {2014}{\natexlab{c}})}\BibitemShut {NoStop}%
\bibitem [{\citenamefont {Moskalets}(2017)}]{moskalets17}%
  \BibitemOpen
  \bibfield  {author} {\bibinfo {author} {\bibfnamefont {M.}~\bibnamefont
  {Moskalets}},\ }\href {\doibase 10.1103/PhysRevB.96.165423} {\bibfield
  {journal} {\bibinfo  {journal} {Phys. Rev. B}\ }\textbf {\bibinfo {volume}
  {96}},\ \bibinfo {pages} {165423} (\bibinfo {year} {2017})}\BibitemShut
  {NoStop}%
\bibitem [{\citenamefont {Ferraro}\ \emph {et~al.}(2010)\citenamefont
  {Ferraro}, \citenamefont {Braggio}, \citenamefont {Magnoli},\ and\
  \citenamefont {Sassetti}}]{ferraro10tunneling}%
  \BibitemOpen
  \bibfield  {author} {\bibinfo {author} {\bibfnamefont {D.}~\bibnamefont
  {Ferraro}}, \bibinfo {author} {\bibfnamefont {A.}~\bibnamefont {Braggio}},
  \bibinfo {author} {\bibfnamefont {N.}~\bibnamefont {Magnoli}}, \ and\
  \bibinfo {author} {\bibfnamefont {M.}~\bibnamefont {Sassetti}},\ }\href
  {\doibase 10.1103/PhysRevB.82.085323} {\bibfield  {journal} {\bibinfo
  {journal} {Phys. Rev. B}\ }\textbf {\bibinfo {volume} {82}},\ \bibinfo
  {pages} {085323} (\bibinfo {year} {2010})}\BibitemShut {NoStop}%
\bibitem [{\citenamefont {Ferraro}\ \emph {et~al.}(2018)\citenamefont
  {Ferraro}, \citenamefont {Ronetti}, \citenamefont {Rech}, \citenamefont
  {Jonckheere}, \citenamefont {Sassetti},\ and\ \citenamefont
  {Martin}}]{ferraro18squeezing}%
  \BibitemOpen
  \bibfield  {author} {\bibinfo {author} {\bibfnamefont {D.}~\bibnamefont
  {Ferraro}}, \bibinfo {author} {\bibfnamefont {F.}~\bibnamefont {Ronetti}},
  \bibinfo {author} {\bibfnamefont {J.}~\bibnamefont {Rech}}, \bibinfo {author}
  {\bibfnamefont {T.}~\bibnamefont {Jonckheere}}, \bibinfo {author}
  {\bibfnamefont {M.}~\bibnamefont {Sassetti}}, \ and\ \bibinfo {author}
  {\bibfnamefont {T.}~\bibnamefont {Martin}},\ }\href {\doibase
  10.1103/PhysRevB.97.155135} {\bibfield  {journal} {\bibinfo  {journal} {Phys.
  Rev. B}\ }\textbf {\bibinfo {volume} {97}},\ \bibinfo {pages} {155135}
  (\bibinfo {year} {2018})}\BibitemShut {NoStop}%
\bibitem [{\citenamefont {Moskalets}(2015)}]{moskalets15}%
  \BibitemOpen
  \bibfield  {author} {\bibinfo {author} {\bibfnamefont {M.}~\bibnamefont
  {Moskalets}},\ }\href {\doibase 10.1103/PhysRevB.91.195431} {\bibfield
  {journal} {\bibinfo  {journal} {Phys. Rev. B}\ }\textbf {\bibinfo {volume}
  {91}},\ \bibinfo {pages} {195431} (\bibinfo {year} {2015})}\BibitemShut
  {NoStop}%
\bibitem [{\citenamefont {Glattli}\ and\ \citenamefont
  {Roulleau}(2016)}]{glattli16wavepackets}%
  \BibitemOpen
  \bibfield  {author} {\bibinfo {author} {\bibfnamefont {D.}~\bibnamefont
  {Glattli}}\ and\ \bibinfo {author} {\bibfnamefont {P.}~\bibnamefont
  {Roulleau}},\ }\href {\doibase https://doi.org/10.1016/j.physe.2015.10.034}
  {\bibfield  {journal} {\bibinfo  {journal} {Physica E: Low-dimensional
  Systems and Nanostructures}\ }\textbf {\bibinfo {volume} {76}},\ \bibinfo
  {pages} {216 } (\bibinfo {year} {2016})}\BibitemShut {NoStop}%
\end{thebibliography}%

\end{document}